\documentclass[aps,prb,twocolumn,amsmath,superscriptaddress,amssymb]{revtex4}
\usepackage{graphicx}
\usepackage{enumitem}
\usepackage{physics}
\usepackage[colorlinks,bookmarks=false,citecolor=red,linkcolor=blue,urlcolor=blue]{hyperref}

\newcommand{\bea}{\begin{eqnarray}}          
\newcommand{\eea}{\end{eqnarray}}

\allowdisplaybreaks

\begin{document}

\title{State selection in frustrated magnets}
\author{Subhankar Khatua}
\email{subhankark@imsc.res.in}
\affiliation{The Institute of Mathematical Sciences, HBNI, C I T Campus, Chennai 600113, India}
\author{Sarvesh Srinivasan}
\affiliation{The Institute of Mathematical Sciences, HBNI, C I T Campus, Chennai 600113, India}
\affiliation{Birla Institute of Technology and Science, Pilani 333031, India}
\author{R. Ganesh}
\email{ganesh@imsc.res.in}
\affiliation{The Institute of Mathematical Sciences, HBNI, C I T Campus, Chennai 600113, India}
\date{\today}

\begin{abstract}
Magnets with frustration often show accidental degeneracies, characterized by a large classical ground-state space (CGSS). Quantum fluctuations may `select' one of these ground states -- a phenomenon labeled `order by (quantum) disorder' in literature. In this article, we examine the mechanism(s) by which such state selection takes place. We argue that a magnet, at low energies, maps to a particle moving on the CGSS. State selection corresponds to localization of the particle at a certain point on this space. We distinguish two mechanisms that can bring about localization. In the first, quantum fluctuations generate a potential on the CGSS space. If the potential has a deep enough minimum, then the particle localizes in its vicinity. We denote this as `order by potential' (ObP). 
In the second scenario, the particle localizes at a self-intersection point due to bound-state formation -- a consequence of geometry and quantum interference. Following recent studies by the present authors, we denote this scenario as `order by singularity' (ObS).  
In either case, localization leads to an energy gap between the ground state(s) and higher-energy states. This pseudo-Goldstone gap behaves differently in the two mechanisms, scaling differently with the spin length. 
We place our discussion within the context of the one-dimensional spin-$S$ Kitaev model. We map out its CGSS which grows systematically with increasing system size. It resembles a network where the number of nodes increases exponentially. In addition, the number of wires that cross at each node also grows exponentially. This self-intersecting structure leads to ObS, with the low-energy physics determined by a small subset of the CGSS, consisting of `Cartesian' states.
A contrasting picture emerges when an additional XY antiferromagnetic coupling is introduced. The CGSS simplifies dramatically, taking the form of a circle. Spin-wave fluctuations generate a potential on this space, giving rise to state selection by ObP under certain conditions. 
Apart from contrasting ObS and ObP, we discuss the possibility of ObS in macroscopic magnets. 
\end{abstract}
%\pacs{}
                                 
\maketitle
\section{Introduction}
The rich physics of frustrated magnetism can often be understood starting from the classical limit, where frustration leads to large `accidental' degeneracies. Unlike a simple bipartite antiferromagnet, frustrated magnets typically allow for a large number of spin configurations that minimize the energy. In such a system, quantum-mechanical fluctuations can play a disproportionately large role in determining the ground state \cite{Shender1982,Rastelli_1987,Henley1989,
Kubo1991,Chubukov1992}. 
A substantial body of literature has developed around this idea, calling it `order by (quantum) disorder' \cite{Chalker2011}. It has also been invoked in materials \cite{Gukasov1988,Bruckel1992,Savary2012}. The term `order by disorder' is also used to denote selection by thermal fluctuations. In this article, we restrict our attention to ground-state selection by quantum fluctuations.

Our goal is to examine the mechanisms by which quantum fluctuations effect ground-state selection. Previous studies have followed a standard prescription which, in our opinion, has not been adequately understood. This prescription is stated as an expansion in powers of $S$ \cite{Shender1996, Henley1989}. The leading term in the Hamiltonian is the $\mathcal{O}(S^2)$ classical energy which may be minimized by multiple classical configurations. In frustrated magnets, such degeneracy is typically `accidental', i.e., it is not related to any symmetry of the Hamiltonian. This allows for selection by quantum effects that emerge at $\mathcal{O}(S)$. They are described by linear spin-wave theory, a framework that \textit{a priori} assumes ordering in a certain classical ground state. Spin-wave modes give rise to zero point energies, taking the form of an $\mathcal{O}(S)$ energy correction. The classical ground-state with the lowest $\mathcal{O}(S)$ correction is deemed to have been `selected' by quantum fluctuations. Although this prescription is widely used, its underpinnings are not well understood. Why is there ordering in a certain classical state in the first place? How sharp is the ordering? What is the regime of validity of this prescription? Does it require a threshold system size and/or a threshold value of $S$? Below, we address such questions by formulating suitable effective low-energy theories.

\begin{figure}
\includegraphics[width=3.3in]{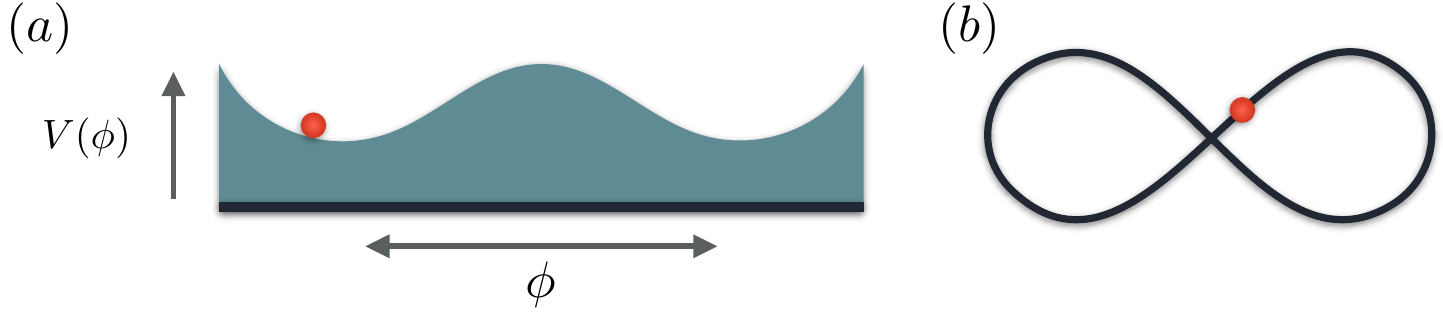}
\caption{
The magnet at low energies maps to a particle moving on the CGSS. State selection is marked by localization of the particle. (a) Order by potential: Quantum fluctuations give rise to a potential on the CGSS. 
The particle localizes at a minimum of the potential. (b) Order by singularity: We have a non-manifold CGSS, exemplified by a figure-of-eight space. A particle moving on this space localizes at the self-intersection point, even in the absence of a potential. 
  }
\label{fig.obdobs}
\end{figure}

To briefly summarize our findings, we describe two distinct selection mechanisms: order by potential (ObP) and order by singularity (ObS). They are depicted as cartoon pictures in Fig.~\ref{fig.obdobs}. 
In general, the low-energy behavior of a magnet maps to a single-particle problem, where the particle moves in the abstract space of classical ground states (CGSS). Selection of a particular ground state corresponds to localization of the particle at some point on this space. In Fig.~\ref{fig.obdobs}(a), we depict localization due to ObP. The particle `sees' a potential that arises from the zero-point energies of quantum fluctuations. If the potential has a sufficiently deep minimum, the particle localizes in its vicinity. This is a generic phenomenon that comes into play wherever accidental degeneracies give rise to a \textit{smooth manifold} as the CGSS.   
In contrast, ObS comes into play when the CGSS self-intersects as shown in Fig.~\ref{fig.obdobs}(b).  Remarkably, in such systems, the particle may localize even in the absence of a potential. It may form a bound state at the self-intersection point, as a consequence of quantum interference and local topology.

A parallel outcome of this article is to provide further support for the notion of ObS.
Previously, ObS has been demonstrated in (i) the XY quadrumer\cite{Khatua2019}, (ii) the Kitaev square\cite{Sarvesh2020} and (iii) the Kitaev tetrahedron\cite{Sarvesh2020}. These are all clusters with four spins. 
However, ObS may also operate in macroscopic magnets with self-intersecting ground-state spaces. A prominent example is the family of pyrochlore magnets\cite{Canals2008,Yan2017}.  
Previous studies on ObS focused on small sizes for practical reasons. Small system size makes it easier to explicitly map out the CGSS and to characterize self-intersections. It also allows for direct evaluation of energy spectra, bringing out features of localization in the low-energy eigenstates. 
In this article, we explore ObS in one-dimensional spin-$S$ Kitaev chains -- a family of models where the system size can be systematically increased. This provides a tunable handle to modify the complexity of self-intersections and the strength of bound-state formation.

\section{The one-dimensional spin-$S$ Kitaev model}
\label{sec.model} 
\begin{figure}
\includegraphics[width=\columnwidth]{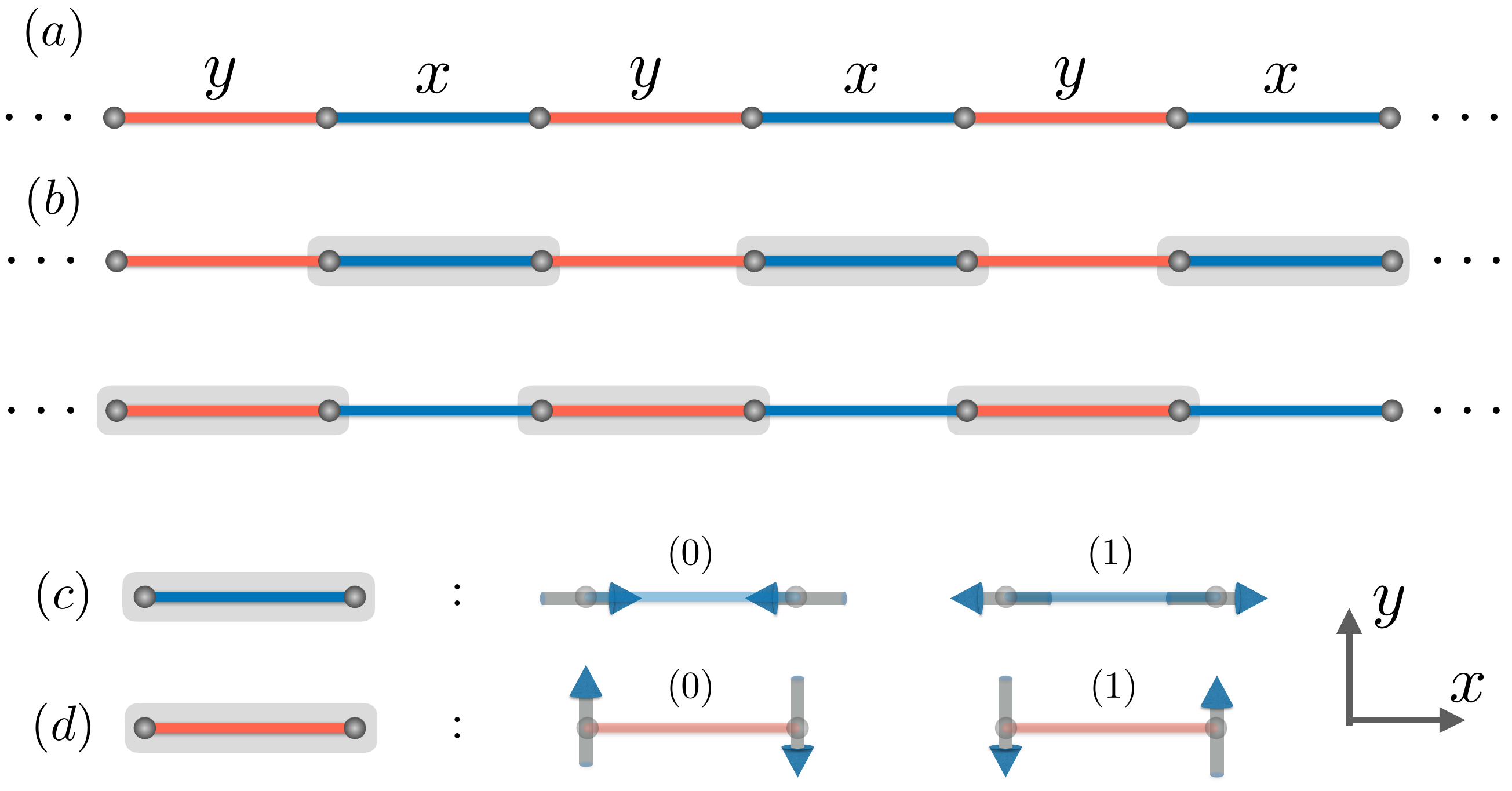}
\caption{The one-dimensional spin-$S$ Kitaev model and its Cartesian states. (a) The one-dimensional chain with alternating $x-x$ and $y-y$ couplings. (b) The two possible dimer covers of this system. One consists of dimers placed on the $x-x$ bonds while the other consists of dimers on the $y-y$ bonds. (c) The two possible ways to orient spins on each $x-x$ dimer to obtain a Cartesian state. One spin is aligned along the $+x$ direction while the other is aligned along $-x$. We label these two configurations on a given dimer as $(0)$ and $(1)$. 
(d) Two possible ways to orient spins on a $y-y$ bond to obtain a Cartesian state.    }
\label{fig.Kitaev_chain}
\end{figure}

We consider the one-dimensional spin-$S$ Kitaev model, first studied by Baskaran, Sen and Shankar (BSS hereafter)\cite{Baskaran2008}. It can also be viewed as a higher spin generalization of the orbital-compass model\cite{Nussinov2009}. It describes a chain of spin-$S$ moments with alternating $x-x$ and $y-y$ couplings, as shown in Fig.~\ref{fig.Kitaev_chain}(a). It is described by the Hamiltonian,
\bea
H_{K} = K \sum_{i} \left[
S_{2i}^x S_{2i+1}^x + S_{2i+1}^y S_{2i+2}^y 
\right].
\label{eq.H_Kitaev}
\eea
Without loss of generality, we assume $K>0$. In a system with $K<0$, its sign can be reversed by a set of local spin rotations at every other site, where the spins are rotated by $\pi$ about the spin-$z$ axis. This model is henceforth referred to as the Kitaev spin chain.

We first consider this model in the $S\rightarrow \infty$ limit, where the spins can be viewed as classical 3-component vectors. The ground states can be found by minimizing the energy with respect to each spin component, using Lagrange multipliers to fix the length of each spin. This approach was first demonstrated by BSS; we recapitulate their arguments in Appendix~\ref{app.CGSS_completeness} for completeness. The minimization procedure leads to the following set of conditions: each pair of neighboring spins must satisfy 
\begin{eqnarray}
S_{i}^{\alpha} = -S_{i+1}^{\alpha}, 
\label{eq.conditions}
\end{eqnarray}
where $\alpha = x$ if $(i,i+1)$ are coupled by an $x-x$ bond or $\alpha=y$ if they are coupled by a $y-y$ bond. It is a non-trivial task to find the set of all configurations that satisfy these conditions. BSS proposed an elegant approach by defining `Cartesian' states and then identifying pathways connecting them. 

Cartesian states are special states that can be immediately seen to satisfy the ground-state conditions. To define a Cartesian state, we start from a dimer cover of the underlying lattice. On each bond that hosts a dimer, we orient the spins so as to minimize the bond energy. In our one-dimensional chain, we have two possible dimer covers as shown in Fig.~\ref{fig.Kitaev_chain}(b). Starting with the dimer cover with dimers on $x-x$ bonds, we anti-align the spins at the ends of each $x-x$ bond. That is, we orient one spin along $\hat{x}$ and the other along $-\hat{x}$. This gives rise to two possible configurations on a given $x-x$ bond as shown in Fig.~\ref{fig.Kitaev_chain}(c). These two configurations can be viewed as two states, $0$ and $1$,  of an Ising variable that lives on the bond.  Proceeding in this manner, we obtain a Cartesian state by independently assigning an Ising variable to each $x-x$ bond. The resulting state immediately satisfies the energy minimization conditions: on each $x-x$ bond, we have $S_{i,x} = -S_{i+1,x}$ by construction. On each $y-y$ bond, $S_{i,y} = -S_{i+1,y}$ is trivially satisfied as $S_{i,y}=S_{i+1,y}=0$. Note that the number of such Cartesian states is exponentially large, corresponding to an extensive number of free Ising spins. The same construction can be carried out starting with the $y-y$ dimer cover. This leads to a family of  Cartesian states with spins pointing along $\pm \hat{y}$. 

BSS next showed that the CGSS contains valleys that connect Cartesian states. Given a pair of Cartesian states that derive from \textit{distinct} dimer covers, there exists a one-parameter family of ground states that smoothly interpolates between them. To visualize this, consider a pair of Cartesian states, one constructed from the $x-x$ dimer cover and the other from the $y-y$ dimer cover. Any such pair of states is smoothly connected by local rotations that are parametrized by a single angle variable, $\phi\in[0,\pi/2]$. For intermediate values of $\phi$, every spin is oriented such that both $x$ and $y$ components take non-zero values, i.e., intermediate states are not Cartesian. Nevertheless, they are also ground states as the energy remains fixed upon tuning $\phi$. Thus, the CGSS can be viewed as a network where the nodes are Cartesian states. The nodes can be grouped into two families: one constructed from the $x-x$ dimer cover and one from the $y-y$ dimer cover. Every node of the x-family connects with every node of the y-family via a one-dimensional pathway. These considerations exhaust all possible ground states, as we argue in Appendix~\ref{app.CGSS_completeness}.

\begin{figure*}
\includegraphics[width=7in]{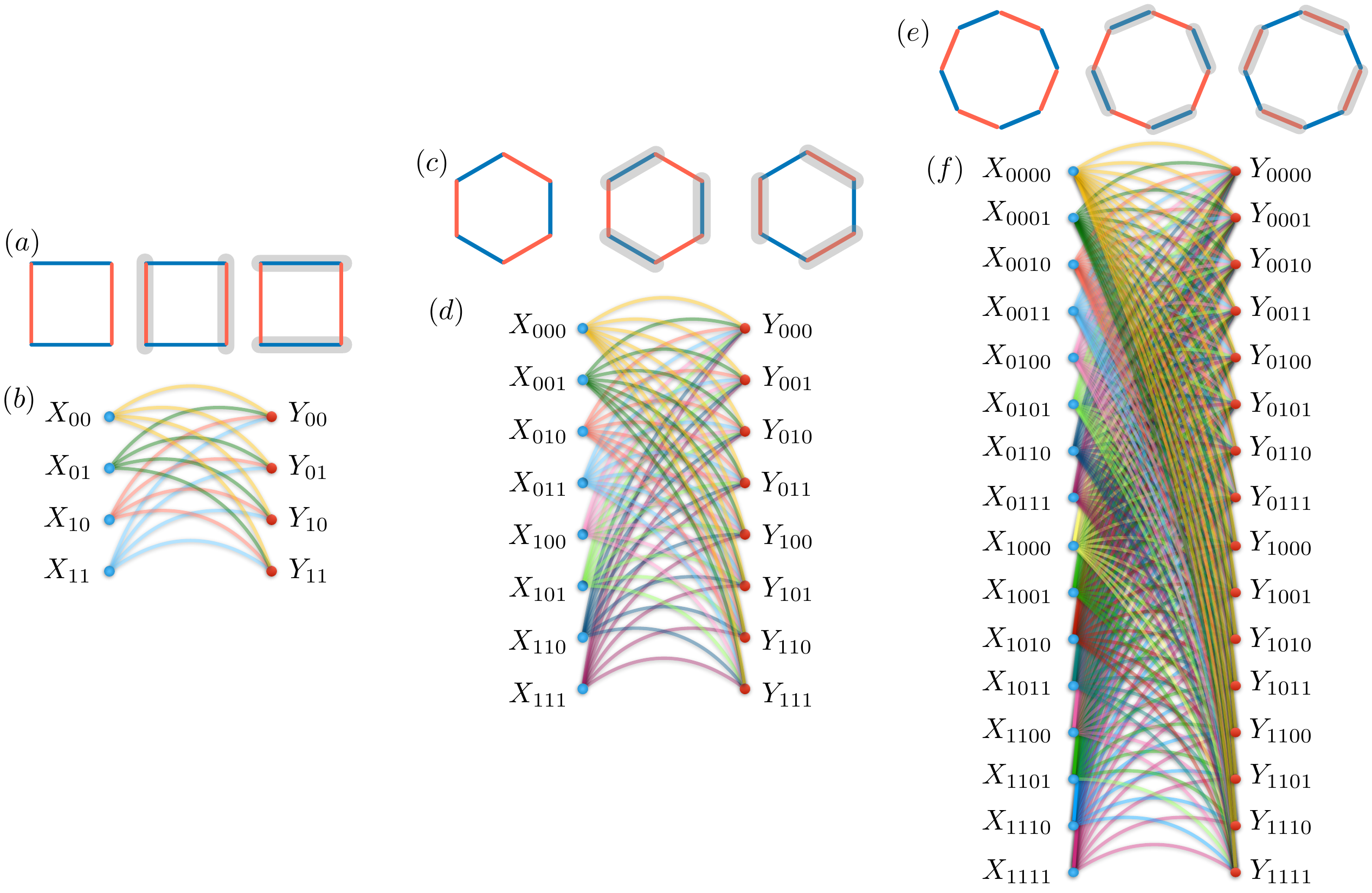}
\caption{The Kitaev spin chain and its CGSS with increasing system size. (a) The Kitaev square, i.e., the chain with $N=4$ spins and periodic boundary conditions. The two possible dimer covers on the square are shown. (b) The CGSS of the $N=4$ problem. We have eight nodes that are divided into two families of four, denoted by $X$ and $Y$. They correspond to Cartesian states derived from the two dimer covers. The subscripts encode spin orientations on each dimer (see text). A node in one family is connected to every node in the other by a one-dimensional path. This represents a smooth one-parameter transformation that connects the two Cartesian states at the end points. (c) The Kitaev hexagon corresponding to the chain with $N=6$ spins and periodic boundary conditions. The two possible dimer covers are shown. (d) The CGSS of the hexagon with sixteen nodes, divided into two families of eight. We have one-dimensional paths connecting each node to every member of the opposite family. (e) The Kitaev octagon corresponding to the chain with $N=8$ spins and periodic boundary conditions. The two possible dimer covers are shown. (d) The CGSS of the octagon with thirty two nodes, divided into two families of sixteen. We have one-dimensional paths connecting each node to every member of the opposite family.     }
\label{fig.wires}
\end{figure*}

We depict the CGSS pictorially in Fig.~\ref{fig.wires}. For concreteness, we take the Kitaev spin chain to consist of an even number of spins, $N$, with periodic boundary conditions. The limit of the infinite chain can be realized by extrapolating to $N\rightarrow\infty$. Fig.~\ref{fig.wires} depicts the CGSS for $N=4,6,8$: the Kitaev square, hexagon, and octagon respectively. In each case, we have two families of nodes, $X$ and $Y$ that derive from the $x-x$ and $y-y$ dimer covers respectively. Within each family, we label every node by a subscript that encodes the Cartesian state as a configuration of Ising moments, following the description given above.

A node in one family is connected to every node in the other via a one-dimensional pathway. The resulting structure of the CGSS is reminiscent of the Lieb-Mattis model for spontaneous symmetry breaking in antiferromagnets\cite{LiebMattis1962}. It describes spins that are grouped into two families. A spin in one family is coupled to every spin in the other. In the same manner, the CGSS here has two families of nodes with pathways connecting all inter-family pairs. 

We note that the CGSS is one-dimensional at generic points. However, it does not have well-defined dimensionality at the nodes which can be viewed as singularities. This indicates that the CGSS is a \textit{non-manifold}. For example, a gradient operator cannot be defined on this space. The non-manifold character increases systematically with system size. For a given $N$, the number of nodes is given by the number of Cartesian states, $N_c = 2\times 2^{N/2} = 2^{N/2+1}$. We have a factor of $2$ for the two possible dimer covers. As each dimer cover has $N/2$ dimers, we have $N/2$ free Ising moments that give rise to $2^{N/2}$ configurations. We also note that the number of `wires' that emanate from a given node is $N_c/2 = 2^{N/2}$, increasing exponentially with system size. This increasing complexity can be seen in Fig.~\ref{fig.wires} for the cases of $N=4,6,8$.

\section{State selection in the Kitaev spin chain: Bound-state formation}
\label{sec.boundstate}
We now discuss state selection in the Kitaev spin chain. 
We have demonstrated that its CGSS is a non-manifold with network-like structure as shown in Fig.~\ref{fig.wires}. Here, we present an effective theory for its low-energy physics.
In previous work by some of us\cite{Khatua2019}, we argued that any magnet, at low energies, maps to the problem of a single particle that is constrained to move on the CGSS. We outlined a proof of this mapping for systems where the CGSS is a smooth manifold, using the spin path integral formalism. We conjectured that the mapping holds for non-manifold cases as well, based on numerical evidence from a few examples\cite{Khatua2019,Sarvesh2020}. 
On the same lines, we argue that the Kitaev spin chain maps to a particle moving on the network-like CGSS. To model its dynamics, we follow Refs.~\onlinecite{Khatua2019,Sarvesh2020} to build a tight binding description. 
 
\subsection{Low-energy physics at self-intersections}
The essential aspect of the problem is the self-intersecting nature of the space at each node. To capture this, we model the vicinity of a single node as a discretized space, shown in Fig.~\ref{fig.Mwires}. We have a central node from which $M$ wires emanate. We label the points as $(j,\gamma)$, where $\gamma=1,\ldots,M$ represents the $M$ wires and $j=1,2,3,\ldots$ is the site index. We label the central node that is common to all wires as $j=0$. We arrive at a tight binding Hamiltonian given by,
\begin{eqnarray}\label{eq.tightbinding_Ham}
H &=& -t \sum_{\gamma=1}^M\sum_{j = 1}^{\infty} \left(c^\dagger_{j,\gamma} c_{j+1, \gamma} + c^\dagger_{j+1,\gamma} c_{j, \gamma}\right)\nonumber\\
&&\hspace{2.5 cm}  -t \sum_{\gamma=1}^M\left( c^\dagger_0 c_{1,\gamma} + c^\dagger_{1,\gamma} c_0 \right).
\end{eqnarray}
This model describes a particle moving on a space of intersecting wires. A traditional Schr\"odinger equation cannot be written down for this problem due to the singular nature of the node. For instance, a kinetic energy operator cannot be defined at the node.

Remarkably, the lowest-energy eigenstate in this system is qualitatively different from others. It represents a bound state that is localized at the singular intersection point. It is described by a simple analytic form, given by
\bea
\psi_{j,\gamma} = A \exp\{ -\alpha_M j\},
\label{eq.psi}
\eea
where $\alpha_M$ represents a decay constant and $A$ is a normalization constant. The wavefunction takes the same form on every wire. To determine the decay constant, we consider a generic site $(j,\gamma)$ that is situated on one of the wires. As it has two neighbors, the eigenvalue equation takes the form,
\bea
E_M  e^{ -\alpha_M j} = -t ( e^{ -\alpha_M (j+1)} + e^{ -\alpha_M (j-1)}), 
\eea
where $E_M$ is the energy eigenvalue. In contrast, at the central node, we have
\bea
E_M = - M t  e^{ -\alpha_M}. 
\eea
From these two equations, we obtain 
\bea
\alpha_M &=& \frac{1}{2} \ln \{ M-1\}, \\
E_M &=& -Mt / \sqrt{M-1}. 
\label{eq.alpha_EM}
\eea
This solution represents a bound state for any $M>2$. We first note that $\alpha_M > 0$ for $M>2$, indicating that the wavefunction decays as we move away from the node. We next consider a neighborhood far from the node, where the system resembles a smooth one-dimensional space. Eigenfunctions that are supported in this region resemble that of a one-dimensional tight binding problem. It follows that their eigenenergies form a continuum between $[-2t,2t]$. The state 
constructed in Eq.~\ref{eq.psi} lies below this continuum as $E_M < -2t$ for any $M>2$. The bound character of the state can be quantified by defining a binding energy, 
\begin{eqnarray}
E_{b,M} = -2t - E_M.
\label{eq.Ebinding_TB}
\end{eqnarray}
We now make an interesting observation regarding this low-energy state and the complexity of the underlying space. If we consider $M$ to be a tunable parameter, we see that the state becomes progressively more bound upon increasing $M$. This can be seen in two quantities that increase monotonically with $M$: the decay constant, $\alpha_M$, and the binding energy, $E_{b,M}$. The parameter $M$, the number of wires that emanates from each node, is a measure of the non-manifold character of the space. 

\subsection{Bound states from quantum interference}
In the preceding paragraphs, we have discussed bound-state formation at a node. We now rationalize this phenomenon at the level of wavefunctions. 
It is well known that a free particle in one dimension behaves like a wave. This can be seen from the wave-like solutions of the Schr\"odinger equation, with $\psi \sim e^{ik x}$. However, the Schr\"odinger equation also allows for exponential solutions, $\psi \sim e^{\pm k x}$. Such solutions are usually ignored due to considerations of normalizability or smoothness. For example, in an infinite wire, such solutions are not normalizable. In a finite wire with periodic conditions, they invariably lead to non-smooth wavefunctions. 
In the case of an open wire with an edge, an exponential solution can be normalizable and smooth. However, it violates the usual boundary condition which demands the wavefunction must vanish at the edge. Unlike these traditional cases, a node-like space as shown in Fig.~\ref{fig.Mwires} provides a rare opportunity. The singular nature of the space at the node removes the need for smoothness. Each wire can host an exponential mode, with the wavefunction decaying with distance from the node. At the node, the wavefunctions on all wires interfere constructively to form a peak. This naturally leads to a localized wavefunction even though there is no potential in the problem.  

The formation of a bound state here bears close similarities with the one-dimensional Schr\"odinger equation with a delta-function potential\cite{Atkinson1975}. The presence of the delta-function (at the origin, let us say) removes the need for smoothness. This allows for a bound state that decays exponentially on either side. This can be viewed as two wires that meet at the origin. They host exponentially decaying modes which interfere constructively to form a peak at the origin. This leads to energy gain from the potential (assumed to be attractive). Effectively, the attractive potential gives rise to a localized ground state. In contrast, at a node where three or more wires meet, the ground state is a bound state even when no potential is involved. The wavefunction can be pictured as follows. The particle sits at the node and simultaneously explores paths that protrude into each of the wires. This allows for enough kinetic energy gain to make this the lowest-energy state. There is no need for an additional potential, unlike the  case of a particle in a delta-function potential.

\subsection{Relevance to the spin-$S$ Kitaev chain}
\begin{figure}
\includegraphics[width=3in]{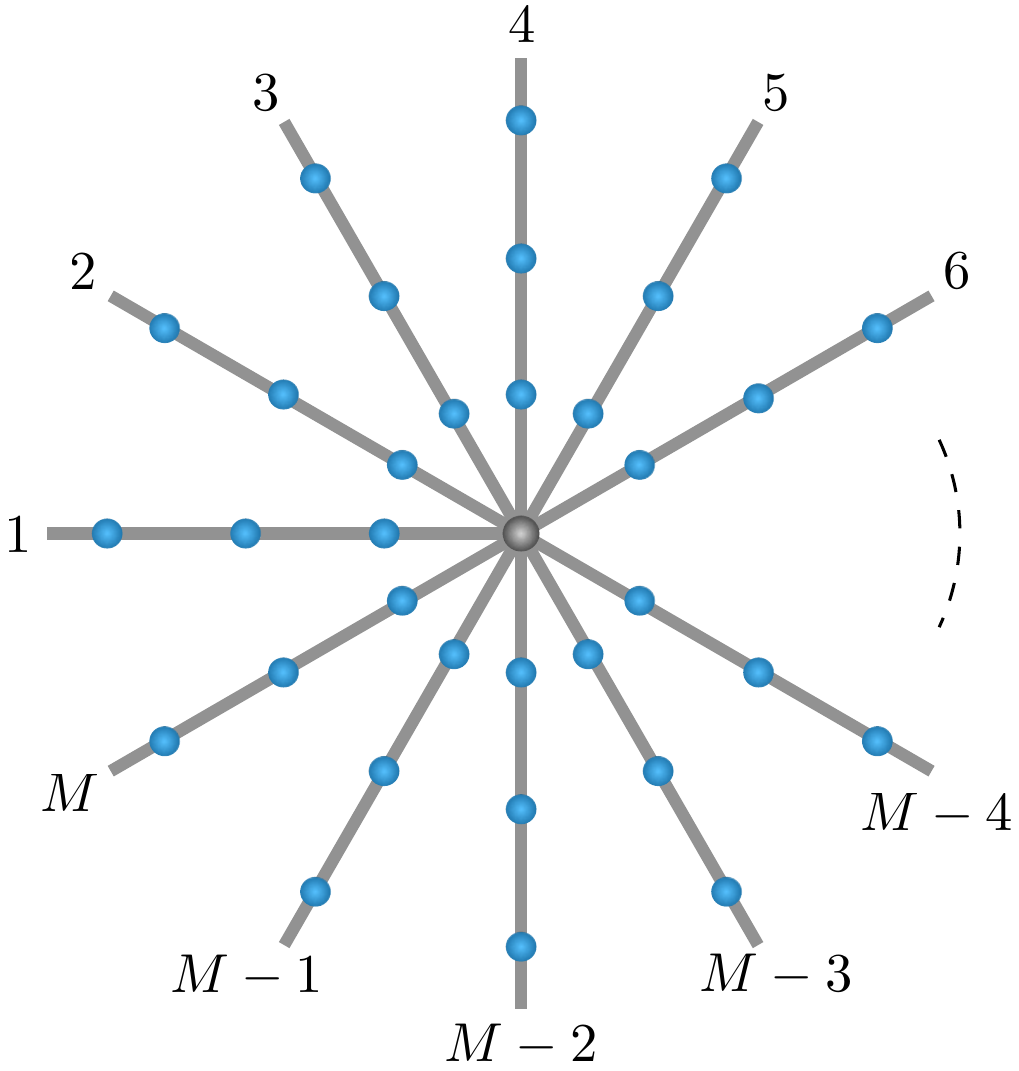}
\caption{
A non-manifold space with $M$ wires intersecting at a point. The space is discretised to allow for a tight binding description. The sites on the legs have two neighbors. The central site is common to all wires and therefore has $M$ neighbors.  
    }
\label{fig.Mwires}
\end{figure}

We now relate this tight binding analysis to the one-dimensional spin-$S$ Kitaev model. We argue that its low-energy physics is described by a particle moving on its non-manifold CGSS. As described in Sec.~\ref{sec.model}, the CGSS consists of nodes and connecting pathways. In the vicinity of each node, the space resembles the tight binding setup pictured in Fig.~\ref{fig.Mwires}. Indeed, there are many such nodes with their number increasing exponentially with system size. 
The analysis in Sec.~\ref{sec.model} clearly brings out the value of $M$, the number of wires emanating from each node. It is given by $M = 2^{N/2}$, where $N$ is the system size. This suggests the following picture. The low-energy spectrum of the Kitaev spin chain consists of bound states whose number grows exponentially with system size. Each of these represents a Cartesian state and small fluctuations in its vicinity. These low-lying states are well separated from other higher-energy states by a `binding energy'. 
Naively, the binding energy grows with increasing system size, scaling as $\sim\sqrt{M} \sim 2^{N/4}$ for large $N$ (see Eq.~\ref{eq.alpha_EM}). This suggests that the binding energy grows without bound as we approach the thermodynamic limit. However, we expect to have an energy cutoff beyond which the particle-on-CGSS picture ceases to describe the magnet. This scale will serve as a natural cutoff for the binding energy. 

We recapitulate that each one-dimensional pathway in the CGSS connects one node to another.  
These wires allow for hybridization among the separate bound states, leading to a spread in the energies of low-lying set. However, the hybridization weakens with increasing system size as the bound states become more tightly bound (as $\alpha_M$ increases with $M$). In the thermodynamic limit, the low-energy physics of the Kitaev spin chain is controlled by a set of bound states that is exponentially large and energetically degenerate.

\section{ObS in exact diagonalization spectra}
\label{sec.ED_Kitaev}
In Sec.~\ref{sec.boundstate}, we have presented an effective description for state selection in the model of Eq.~\ref{eq.H_Kitaev}. We have argued that the low-energy physics is determined by bound states at the nodes of the CGSS. We now support this assertion with evidence from exact diagonalization spectra. 

\subsection{Methodology}
\label{ssec.ed_methodology}
We take the Kitaev spin chain to consist of $N$ sites with periodic boundaries. We present results for the cases of $N=4,6,8,10$ and for various values of $S$, the spin length. The Hilbert space is given by $(2S+1)^N$, growing rapidly with system size. In order to study the spectrum, we use the following two symmetries of the problem: (a) The system is invariant under a global spin rotation by $\pi$ about the spin-z axis. This symmetry allows us to divide the Hilbert space into odd and even magnetization sectors. (b) The Hamiltonian is invariant under a combination of unit-translation and a $\pi/2$ rotation about the spin-z axis. This allows us to define quasi-momentum blocks that are independent. These two symmetries were used to study the $N=4$ problem in Ref.~\onlinecite{Sarvesh2020}. For $N>4$, despite these symmetries, block sizes are prohibitively large for full diagonalization. We use Lanczos diagonalization implemented using the ARPACKPP package \cite{arpackpp}, focusing on the lowest few eigenstates. We take advantage of the sparse nature of the matrices, which in turn is a consequence of the local nature of terms in the Hamiltonian.

\subsection{Bound states in the spectra}
A detailed study of the Kitaev spin chain with $N=4$ has been presented in Ref.~\onlinecite{Sarvesh2020}. Before moving on to larger system sizes, we recapitulate the key features of the $N=4$ case. Its low-energy spectrum is characterized by eight bound states. Note that this is precisely the number of CGSS nodes or Cartesian states ($N_c=2^{N/2 +1} = 8$ for $N=4$). These states have strong overlaps with (quantum analogues of the) Cartesian states.  Energetically, they are separated from higher-energy states by a `binding energy' that increases linearly with $S$.  

\begin{figure*}
\includegraphics[width=6.5in]{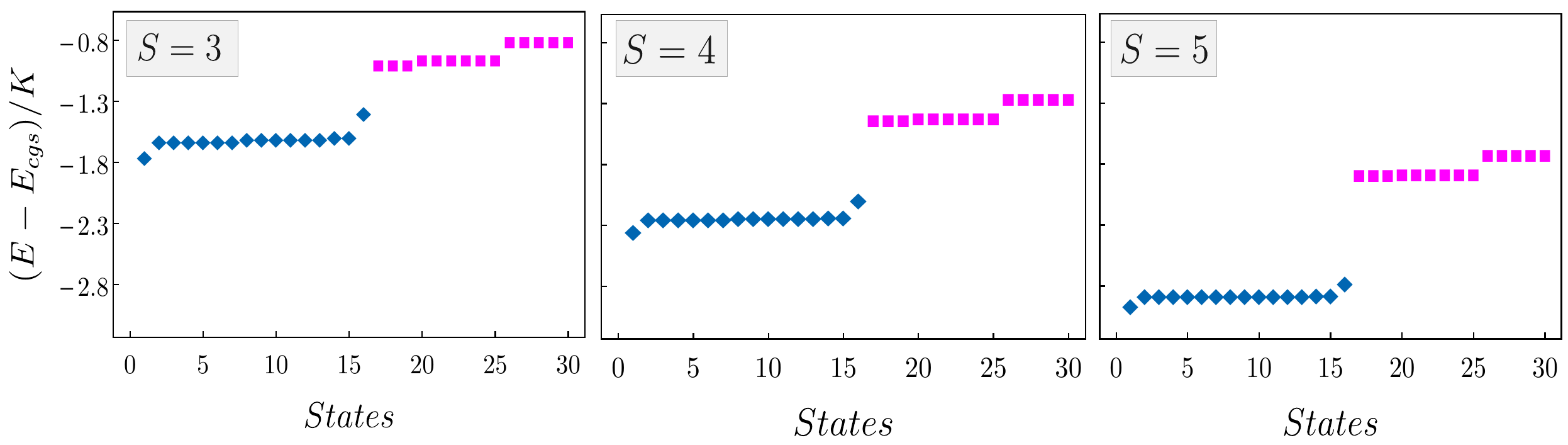}
\caption{Low-energy spectrum of the Kitaev spin chain with $N=6$.}
\label{fig.N6}
\end{figure*}

\begin{figure*}
\includegraphics[width=5in]{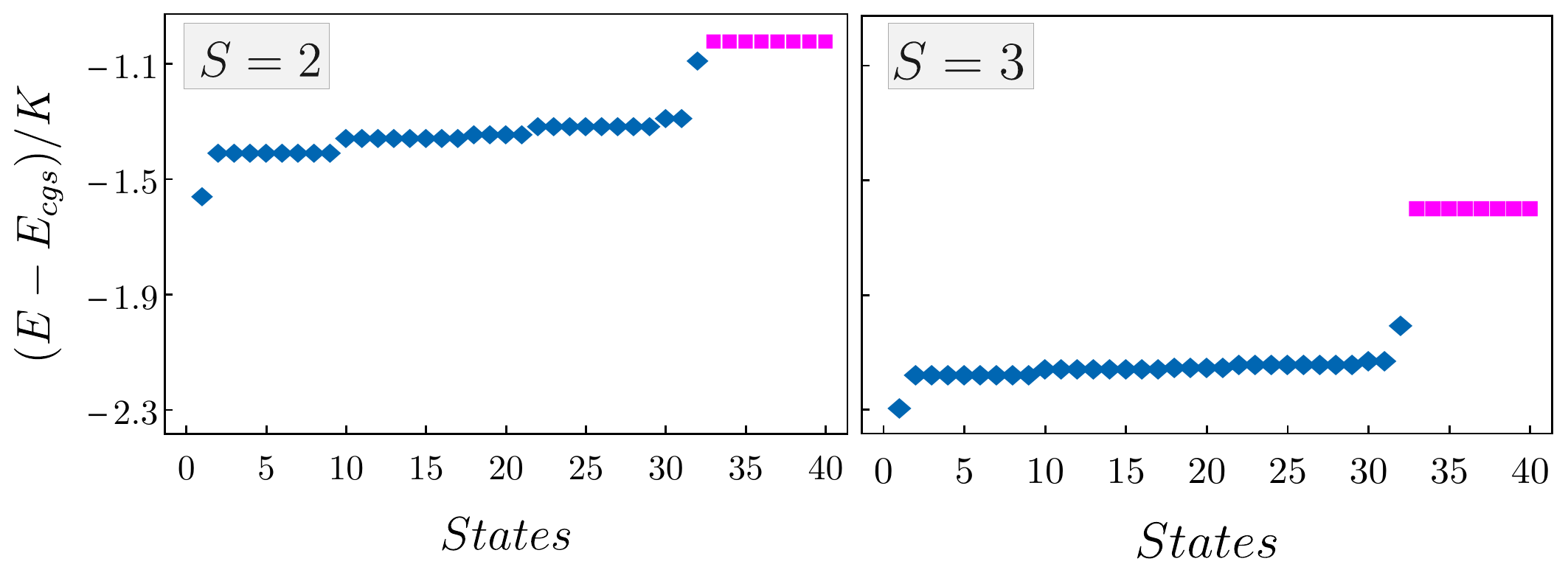}
\caption{Low-energy spectrum of the Kitaev spin chain with $N=8$.}
\label{fig.N8}
\end{figure*}

We now present numerical spectra for $N>4$ which also show clear evidence of bound-state formation. In Fig.~\ref{fig.N6}, we show the spectra for $N=6$ and $S=3,4,5$. We indicate the lowest 16 states using a different colour to show that they are energetically separated from other states. From the analysis in Sec.~\ref{sec.boundstate}, we indeed expect $N_c = 2^{N/2 + 1} = 16$ bound states for $N=6$. 
We next show the low-energy spectra for $N=8$ and $S = 2, 3$ in Fig.~\ref{fig.N8}. Here, we see 32 bound states -- in agreement with the number of Cartesian states for $N=8$. 

\begin{figure}
\includegraphics[width=3in]{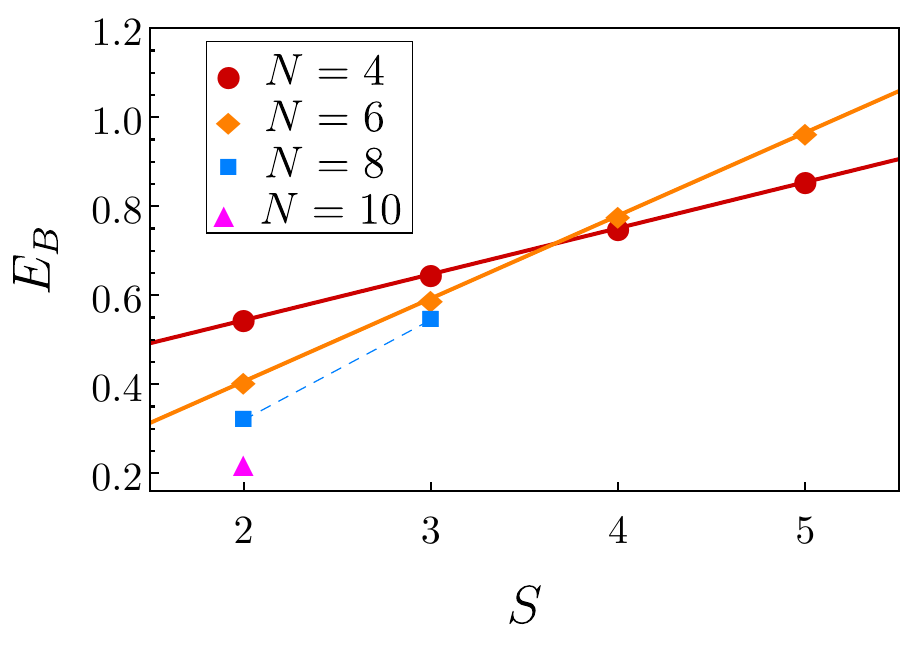}
\caption{Binding energy vs. $S$ for $N=4,6,8$ and $10$. }
\label{fig.BEvsS}
\end{figure}

As seen for $N=6$ in Fig.~\ref{fig.N6}, the energy gap that separates bound states from others increases with increasing $S$. To quantify this, we define $\bar{E}_{1-16}$ as the mean of the energies of the lowest sixteen states. We define the binding energy as $E_{b,N=6} = E_{17} - \bar{E}_{1-16}$, where $E_{17}$ is the energy of the seventeenth state, i.e., the lowest unbound state. In the same manner, we define the binding energy for $N=8$ as $E_{b,N=8} = E_{33} - \bar{E}_{1-32}$. In Fig.~\ref{fig.BEvsS}, we show the variation of binding energy with $S$. We have included data from Ref.~\onlinecite{Sarvesh2020} for $N=4$. 
Note that for $N=8$, we only have data for $S=2$ and $3$. We are unable to access higher values of $S$ as the Hilbert space dimension is too large. We have also added the binding energy for $N=10$ where the binding energy is defined as $E_{b,N=10} = E_{65} - \bar{E}_{1-64}$. We only have binding energy for $S=2$ as higher spin values are not accessible. For all $N$, we only show the binding energy for $S \geq 2$ as we do not find clear energy separation with $S=1$.
 
Fig.~\ref{fig.BEvsS} shows a clear linear rise in the binding energy with $S$. This indicates that the binding energy grows linearly with $S$ for any value of $N$. This scaling relation is potentially a signature of ObS, indicating strengthening of selection with increasing $S$. We will contrast this relation with the case of ObP in Sec.~\ref{sec.HKJ} below.
 
\subsection{Approaching the thermodynamic limit}
\label{ssec.approaching}
We now try to extrapolate the observed physics to the infinite Kitaev spin chain. Does bound-state formation occur even as $N\rightarrow \infty$? To examine this issue, we first consider Fig.~\ref{fig.BEvsS}. This plot presents the variation of binding energy with $S$ when $N$ is held fixed. We find a linear increase, with the binding energies falling along a straight line.  
In addition, the slope of the binding energy vs. $S$ curve increases with increasing $N$.

\begin{figure}
\includegraphics[width=3.5in]{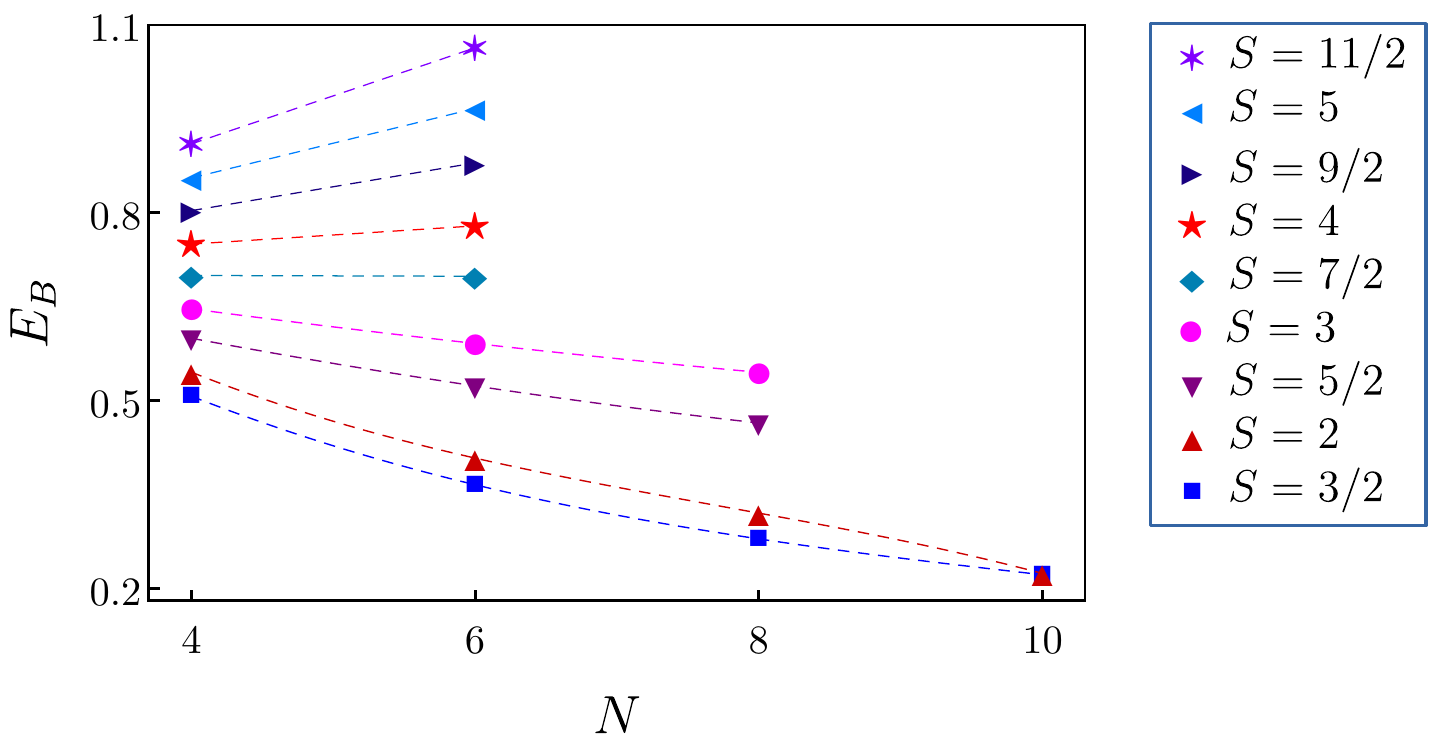}
\caption{Binding energy for various system sizes and $S$. We include data for all systems that are accessible within our numerical limitations. For $S \geq 3.5$, we can only access $N=4$ and $6$. For large $S$ values, we see that the binding energies increase with system size. However, for $S < 3.5$, the binding energy decreases with system size. The dashed lines are guides to the eye. }
\label{fig.BEvsNS}
\end{figure}

To further explore whether ObS survives in the thermodynamic limit, we focus on the behavior of the binding energy as $N\rightarrow \infty$. In Fig.~\ref{fig.BEvsNS}, we plot the binding energies for all cases that are accessible within our numerical constraints. For all cases shown here, we find the correct number of bound states. That is, we find a gap that separates the lowest $2^{N/2 + 1}$ states from the higher states. We define the binding energy in the same manner as for $N=6,8$ described above. 
We only have binding energy data for a limited range of $(N,S)$ values where the Hilbert space size remains manageable. This data range is not sufficient to deduce the functional dependence of binding energy on system size. Nevertheless, we list some observations that are consistent with the data. When $S$ is kept fixed at a small value, binding energy decreases with increasing $N$. Instead, if $S$ is kept fixed at a large value, binding energy increases with $N$. This suggests that binding energy changes non-monotonically outside the accessible region. For instance, it is possible that for large $S$ values, binding energy may eventually decrease beyond a threshold system size.
To rationalize this non-monotonic behavior, we argue that the behavior of binding energy vs. $N$ is determined by two conflicting effects. At low energies, the system maps to the particle-on-CGSS picture. Within this picture, binding energy increases rapidly with $N$ as given by Eq.~\ref{eq.alpha_EM}. However, the mapping to the particle-on-CGSS picture is valid at low energies -- below a certain cutoff energy scale. In Ref.~\onlinecite{Khatua2019}, for the case of a smooth manifold CGSS, it was shown that the particle-on-CGSS picture emerges upon integrating out `hard' modes. This suggests that the particle-on-CGSS picture generally holds below the energy of the lowest hard mode. In the case of the Kitaev spin chain, on general grounds, we expect the energy of the lowest hard mode to decrease with increasing system size. This suggests that the cutoff decreases with increasing $N$. The increasing `bare' binding energy and the decreasing cutoff together determine the observed binding energy.

A precise understanding of the ObS binding energy in the thermodynamic limit requires a careful analysis on the lines of Ref.~\onlinecite{Rau2019}. Our limited numerical results do not lead to a definite conclusion.  
We content ourselves with the following observation based on our data. 
The binding energy scales linearly with $S$, providing a hallmark of ObS. This holds true for $N=4,6,8$ and possibly beyond. Notably, the slope increases with increasing $N$, indicating that ObS strengthens as $N\rightarrow \infty$.

\section{Regularizing the CGSS to remove singularities: the $K$-$J$ model}
\begin{figure}
\includegraphics[width=3in]{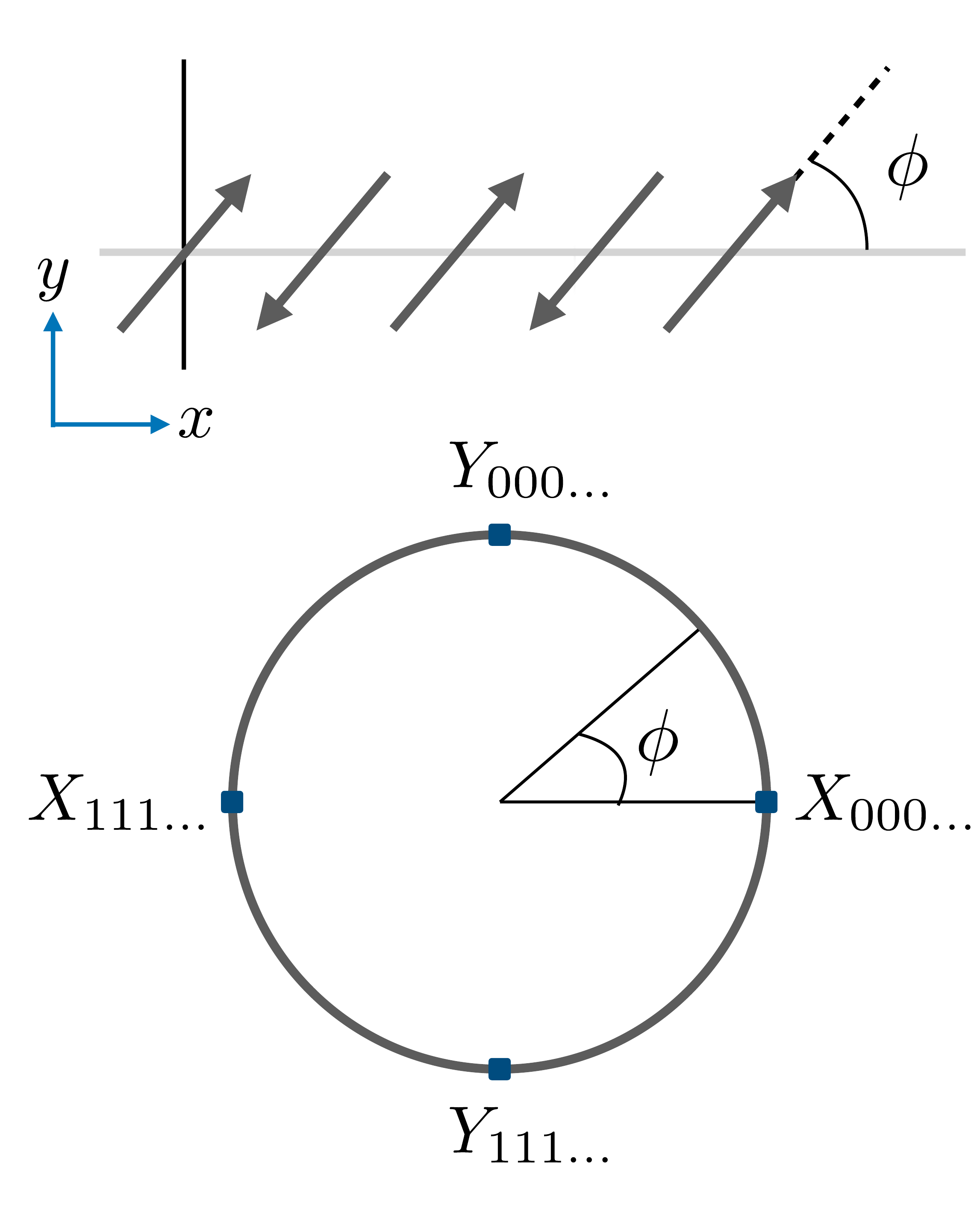}
\caption{Top: Classical ground-state of the K-J model, parametrized by an angle $\phi$. 
Bottom: The CGSS forming a circle. We indicate four points on the circle that correspond to Cartesian states as described in the context of the Kitaev spin chain. 
}
\label{fig.KJCGSS}
\end{figure}

We have established that the model of Eq.~\ref{eq.H_Kitaev} has a non-manifold CGSS. This allows for ObS as we have demonstrated in Secs.~\ref{sec.boundstate} and \ref{sec.ED_Kitaev}. We now introduce a second system where ObS cannot arise, but ObP can. To do this, we introduce an additional coupling that `regularizes' the CGSS of the Kitaev spin chain, 
\bea
H_{K-J} = H_K + J \sum_{i} \left[
S_{i}^x S_{i+1}^x + S_{i}^y S_{i+1}^y 
\right].
\label{eq.HKJ}
\eea
We have introduced an XY antiferromagnetic coupling with strength $J$ ($J>0$). This coupling selects a particularly simple subset of the Kitaev CGSS, consisting of states of the form $\vec{S}_i = (-1)^i\{\cos\phi ~ \hat{x} + \sin\phi ~ \hat{y}\}$. These states are immediately seen to minimize the antiferromagnetic ($J$) part of the Hamiltonian. At the same time, they satisfy the conditions encoded in Eq.~\ref{eq.conditions} for minimizing the Kitaev term. 
As this family of states is parametrized by a single angle variable, the CGSS of $H_{K-J}$ is equivalent to a circle. We depict the CGSS pictorially in Fig.~\ref{fig.KJCGSS}. The CGSS here represents \textit{accidental} degeneracy, as there is no Hamiltonian symmetry that relates states corresponding to different values of $\phi$. In the figure, we indicate four special points on the CGSS that correspond to $\phi = 0,\pi/2,\pi,3\pi/2$. These four states are Cartesian states as defined in the context of the Kitaev spin chain in Sec.~\ref{sec.model}. Thus, the CGSS of the $K$-$J$ model can be viewed as consisting of four `nodes' that are connected by one-dimensional pathways. This further illustrates that the CGSS of the $K$-$J$ model is a subset of that of the Kitaev spin chain. 

Crucially, this new CGSS is a smooth manifold with no singularities. The physics of the Kitaev spin chain is recovered in the limit of $J\rightarrow 0$. Conversely, this model reduces to the standard XY antiferromagnet as $J\rightarrow \infty$. In the latter limit, we recover rotational symmetry that protects the degeneracy. For any finite value of $J$, the degeneracy is accidental, allowing for the possibility of state selection. 

\section{State selection in the $K$-$J$ model}
\label{sec.HKJ}
As the CGSS of the $K$-$J$ model is a smooth manifold, it does not allow for bound-state formation of the form described in Sec.~\ref{sec.boundstate} above. State selection here requires a different mechanism -- that of a potential superimposed on this space. The ObP paradigm proposes that such a potential is generated by quantum fluctuations. We discuss this scenario here. 

We present a standard order-by-quantum-disorder calculation, valid in the semiclassical large-$S$ limit.  
We invoke quantum fluctuations in the form of spin waves using the Holstein-Primakoff prescription \cite{Holstein1940}. They give rise to a zero point energy contribution that modifies the ground-state energy. This contribution breaks the accidental degeneracy to `select' certain ground states. Details regarding the spin-wave calculation are given in Appendix~\ref{app.sw_JK}. In Fig.~\ref{fig.JK_deltaE}, we plot the zero point energy contribution along the CGSS, i.e., zero point energy vs. $\phi$. As expected, this correction to the energy breaks the degeneracy of the CGSS. It has minima at four distinct values of $\phi$. These four correspond to alternating spins along the $x$ or $y$ directions. In fact, they are Cartesian states in the language of the pure Kitaev spin chain. 

\begin{figure}
\includegraphics[width=\columnwidth]{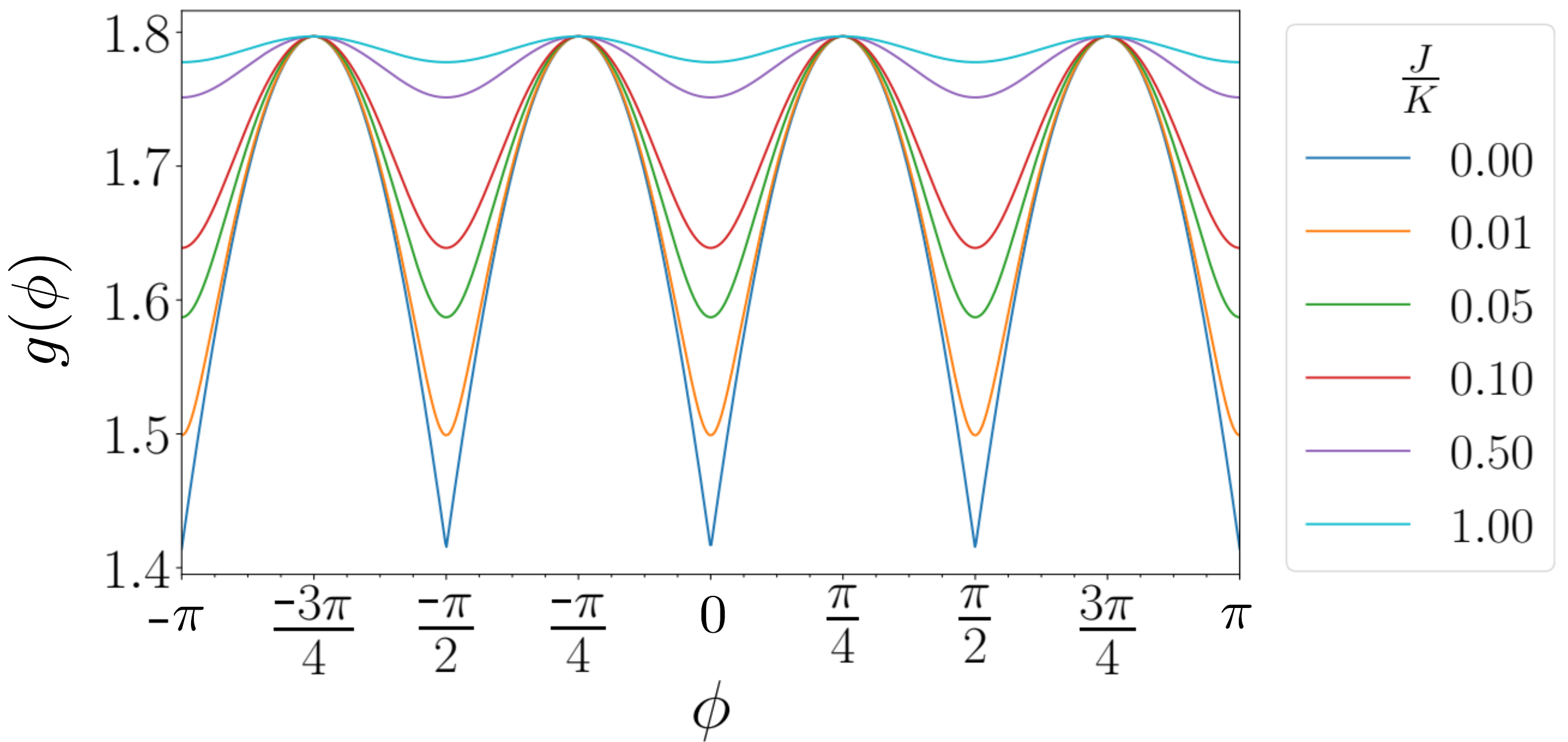}
\caption{Zero point energy in units of $(2J+K)S$ vs. $\phi$ for various values of $J/K$. We find four minima, with the minima becoming deeper with decreasing $J/K$.}
\label{fig.JK_deltaE}
\end{figure}

A physical picture for ObP emerges from the analyis of Rau \textit{et al.} in Ref.~\onlinecite{Rau2019}. We present this picture in somewhat different language here. The low-energy physics of the K-J model can be understood by constructing a non-linear sigma model using the following parametrization,
\begin{eqnarray}
\mathbf{S}_i (\phi, m)= \left\{ \begin{array}{c}
S \{ \hat{n}_\phi + m \hat{z}\}, ~i = 2n \\
S \{ -\hat{n}_\phi + m \hat{z}\}, ~i = 2n+1 \\
\end{array} \right.,
\end{eqnarray}
where $\hat{n}_\phi = \{ \cos \phi, \sin\phi, 0\}$. To preserve normalization, we assume $m \ll 1$. We have ignored other independent contributions that can enter in the spin configuration (e.g., a staggered magnetization along $\hat{z}$) as these degrees of freedom can be integrated out safely. In other words, we have retained the `soft' degree of freedom and its conjugate mode. The angle $\phi$ is `soft' as it does not change the energy. The magnetization $m$ is canonically conjugate to $\phi$. This can be seen from the form of the Lagrangian in the path-integral partition function,
\begin{eqnarray}
\nonumber \mathcal{L} (m,\phi,\dot{\phi}) &=& -iNS m \dot{\phi} + (aN S^2) m^2 +  N S g(\phi) \\
&=& (-iNSm) \dot{\phi} + \frac{(NSm)^2}{(N/a)}  +  NSg(\phi) .
\label{eq.Lagrangian}
\end{eqnarray}
It is written following the imaginary-time convention. Here, $-iNSm \dot{\phi}$ represents the Berry phase. This is a geometric contribution that arises from the area swept out by each spin vector. The other terms in the action constitute the Hamiltonian or the energy. Along the $m$ direction, the energy increases quadratically with a stiffness coefficient denoted as $a$. This contribution is proportional to $S^2$, as $m$ takes us away from the classical ground-state space. It is also proportional to $N$ as the energy cost scales linearly with system size. In the soft direction, we assume a potential, $g(\phi)$. This term is allowed on symmetry grounds since $\phi$ represents an accidental degeneracy. 

Rau \textit{et al.} provides a simple prescription to obtain the form of $g(\phi)$ to $\mathcal{O}(S)$. This prescription has also been used in earlier studies as a heuristic to understand state selection\cite{Shender1982,Belorizky1980}. The prescription dictates that the potential at each point on the CGSS, $g(\phi)$, should be taken to be the sum of zero point energies of all spin-wave modes, 
\begin{eqnarray}
g(\phi) \equiv \frac{1}{N} \sum_{\mathbf{k},\alpha} \epsilon_\phi (\mathbf{k},\alpha).
\label{eq.gphi}
\end{eqnarray}
Here, $\epsilon_\phi (\mathbf{k},\alpha)$ represents the eigenenergies or frequencies of spin-wave modes (see Appendix~\ref{app.sw_JK} for explicit expressions). Each spin-wave mode is characterized by two quantum numbers: momentum $\mathbf{k}$ and an internal index (or a band index) $\alpha$. We divide the zero point energies by system size $N$ in order to obtain the intensive energy. The spin-wave energies scale linearly in $S$; we have taken this factor of $S$ out in Eq.~\ref{eq.Lagrangian}.

From the form of the Lagrangian in Eq.~\ref{eq.Lagrangian}, we draw an analogy to single-particle quantum mechanics. We identify
\begin{eqnarray}
\nonumber \phi &\leftrightarrow &  q, ~~
(NSm) \leftrightarrow  p,\\
\{ NSg(\phi) \} &\leftrightarrow & V(q) ,~~(N/a) \leftrightarrow 2\mu.
\label{eq.variables}
\end{eqnarray}
With this identification, the Lagrangian resembles that of a particle moving in one dimension with 
$\mathcal{L} \sim -ip\dot{q} + \frac{p^2}{2\mu} + V(q)$. The position and momentum coordinates are given by $q$ and $p$ respectively, with the potential given by $V(q)$. The mass of the particle is denoted as $\mu$. 

\subsection{Localization in a potential well} 
\label{ssec.well}
\begin{figure}
\includegraphics[width=\columnwidth]{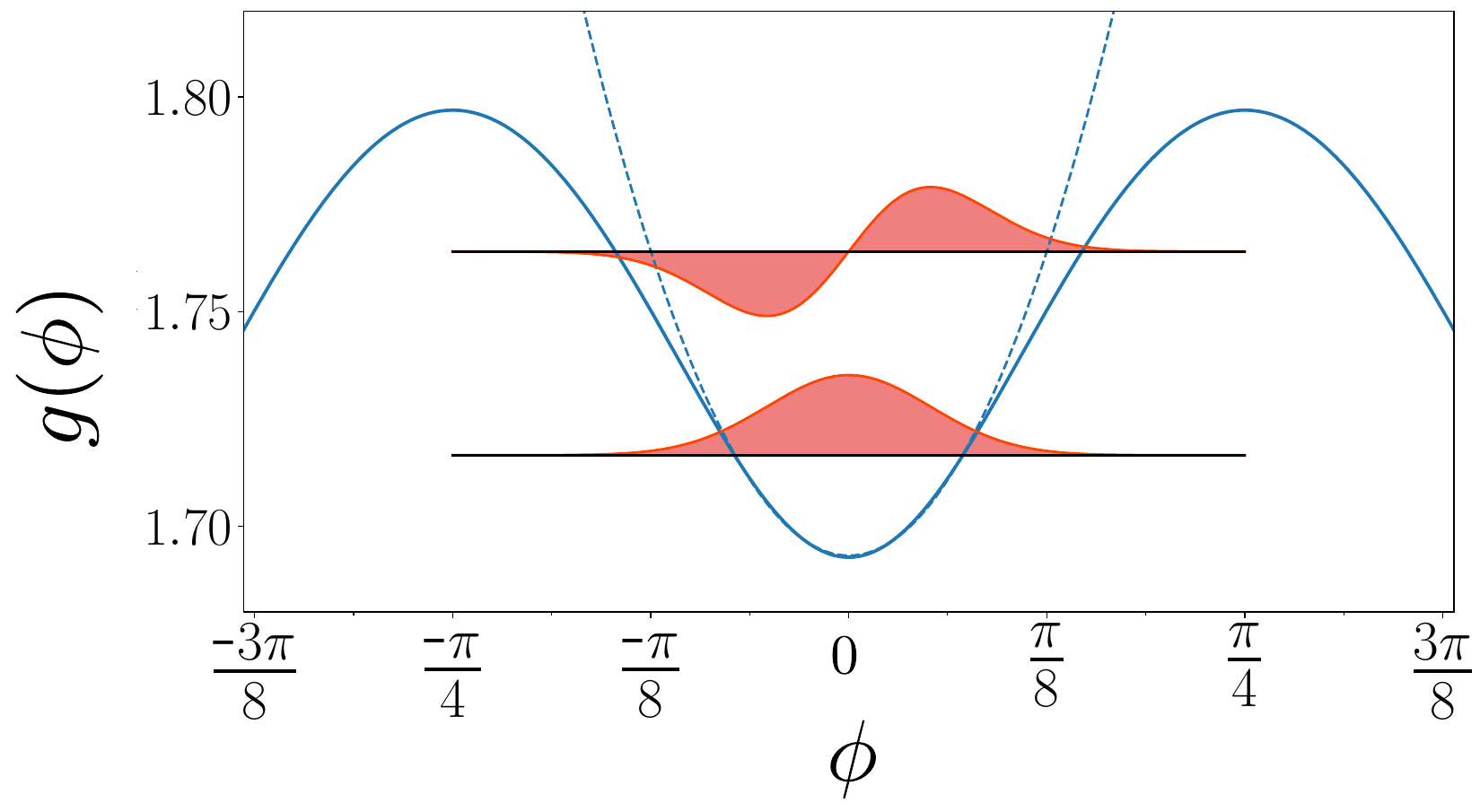}
\caption{ObP as localization due to potential. The potential stems from zero point energy, shown here for $J/K = 0.2$. Here, $g(\phi)$ has been taken in units of $(2J+K)S$. At low energies, the particle stays in the vicinity of a minimum where the potential resembles that of a harmonic oscillator. The particle localizes by settling in the ground state. The gap to the first excited state is the pseudo-Goldstone gap. }
\label{fig.harmonic}
\end{figure}

We now follow Eq.~\ref{eq.gphi} to interpret the zero point energy contribution plotted in Fig.~\ref{fig.JK_deltaE} as the potential seen by the particle. Clearly, the potential has four minima at $\phi = 0, \pi/2, \pi,$ and $3\pi/2$. In the vicinity of these points, the potential resembles a harmonic well (as long as $J\neq 0$). If this well is deep enough, it will localize the particle and lead to state selection. The lowest-energy states of the system will then resemble the eigenstates of a harmonic oscillator as shown in Fig.~\ref{fig.harmonic}. The analysis of Rau \textit{et al.} assumes this scenario and evaluates the gap to the first excited state. It arrives at a scaling relation where the gap scales as $S^{1/2}$. Here, we restrict our attention to type-I systems in the language of Rau \textit{et al.} as our $K$-$J$ model falls in this class. Below, we expand their arguments to examine when localization occurs. For example, is there a threshold system size below which there is no localization? 

Can the potential localize the particle? This is a question of competition between kinetic energy (in a delocalized state) and potential energy (in a localized state). We quantify this notion by comparing two suitably defined energy scales. To quantify the potential energy gain due to localization, we first define an energy $E_1$ as follows. We assume that the potential has a deep minimum at $\phi = \phi_0$. In its vicinity, the potential takes the form $V(\phi_0 + \delta \phi) \sim V_0 + NS \gamma (\delta \phi)^2$ where $\gamma$ is a proportionality constant. Here, the problem resembles a simple harmonic oscillator. If the particle is restricted to this region, it resides in eigenstates of the harmonic oscillator problem. The lowest-energy state is a Gaussian localized around $\phi_0$. Higher-energy states are progressively broader. They are separated by a characteristic energy spacing, $E_1 \equiv \hbar \omega \sim \sqrt{NS \gamma / (N/a)} \sim \sqrt{a\gamma S}$. This scale quantifies the gain in potential energy when the particle localizes in the ground state rather than in excited states. In fact, it is precisely this scale that sets the pseudo-Goldstone gap in the analysis of Rau \textit{et al.}

We next define $E_2$ to quantify the kinetic energy of a delocalized state. To do so, we ignore the potential for the moment, assuming that the system corresponds to a free particle. The particle lives on a ring as the position coordinate satisfies periodic boundary conditions with $\phi \equiv \phi \pm 2\pi$. The stationary states of this particle are given by plane waves, $\psi \sim e^{i p \phi}$,  with energies $\epsilon_p = \frac{p^2}{2\mu} \equiv \frac{a}{N} p^2$, where $p$ is the momentum eigenvalue. Periodic boundaries constrain the allowed momentum values to $p=0,\pm1,\pm2,\ldots$. The resulting energies are characterized by a scale $E_2 \equiv \frac{a}{N}$. 

We now argue that localization requires $E_1 \gg E_2$, i.e., the energy scale of the potential energy must exceed that of the kinetic energy. This leads to 
\begin{eqnarray}
\sqrt{a \gamma S} \gg \frac{a}{N} \implies N \sqrt{S} \gg \sqrt{\frac{a}{\gamma}}.
\label{eq.criterion}
\end{eqnarray}
This serves as a criterion for state selection by ObP. We require sufficiently large values of the system size, $N$, and the spin quantum number, $S$ so that $(N\sqrt{S})$ exceeds a threshold value. As an application of these ideas, we present exact diagonalization results on the $K$-$J$ model in the following section.

We make one final observation here regarding state selection in the $J\rightarrow 0$ limit. The physics of the $K$-$J$ model with $J>0$ does not connect smoothly to the pure Kitaev limit. As long as $J>0$, the potential in Fig.~\ref{fig.JK_deltaE} has four harmonic wells. We can understand localization by considering $(N\sqrt{S})$, the quantity in Eq.~\ref{eq.criterion}, as a tuning parameter. When this quantity is small, the physics corresponds to that of a free particle. The potential $g(\phi)$ in Eq.~\ref{eq.Lagrangian} remains insignificant. However, as this quantity increases, the effect of the potential becomes stronger. We approach the case of a particle localized in a harmonic well. We now consider the effect of decreasing $J$ as shown in Fig.~\ref{fig.JK_deltaE}. The curvature of each well increases steadily. At $J =0$, the potential becomes singular, resembling the $\vert x \vert$ function rather than a harmonic well. 
However, a bigger change occurs at this point. As argued in Sec.~\ref{sec.model}, the underlying space (the CGSS) becomes larger with multiple additional pathways. In this limit, the free-particle problem already shows localization due to bound-state formation as described in Sec.~\ref{sec.boundstate}. Quantum fluctuations can generate an additional potential which will take the same form on all pathways. We may surmise that this potential is of the $\vert x \vert$ type. Even if the potential is strong, it only serves to make the bound states even more bound. This sets ObS at $J=0$ apart from ObP at $J>0$. While ObP can be made weaker by changing a tuning parameter, ObS cannot. 

\section{ObP in exact diagonalization spectra}

We have established that the $K$-$J$ model of Eq.~\ref{eq.HKJ} has a circle as its CGSS. We have argued that at low energies, the model maps to a particle moving on a circle. We have further argued for a symmetry-allowed potential that localizes the particle as long as $N\sqrt{S}$ is sufficiently large. We now present evidence for these statements in the form of exact diagonalization results. We follow the methodology described in Sec.~\ref{ssec.ed_methodology} in the context of the pure Kitaev spin chain. The symmetries described there hold for the $K$-$J$ model as well. 

\begin{figure}
\includegraphics[width=3in]{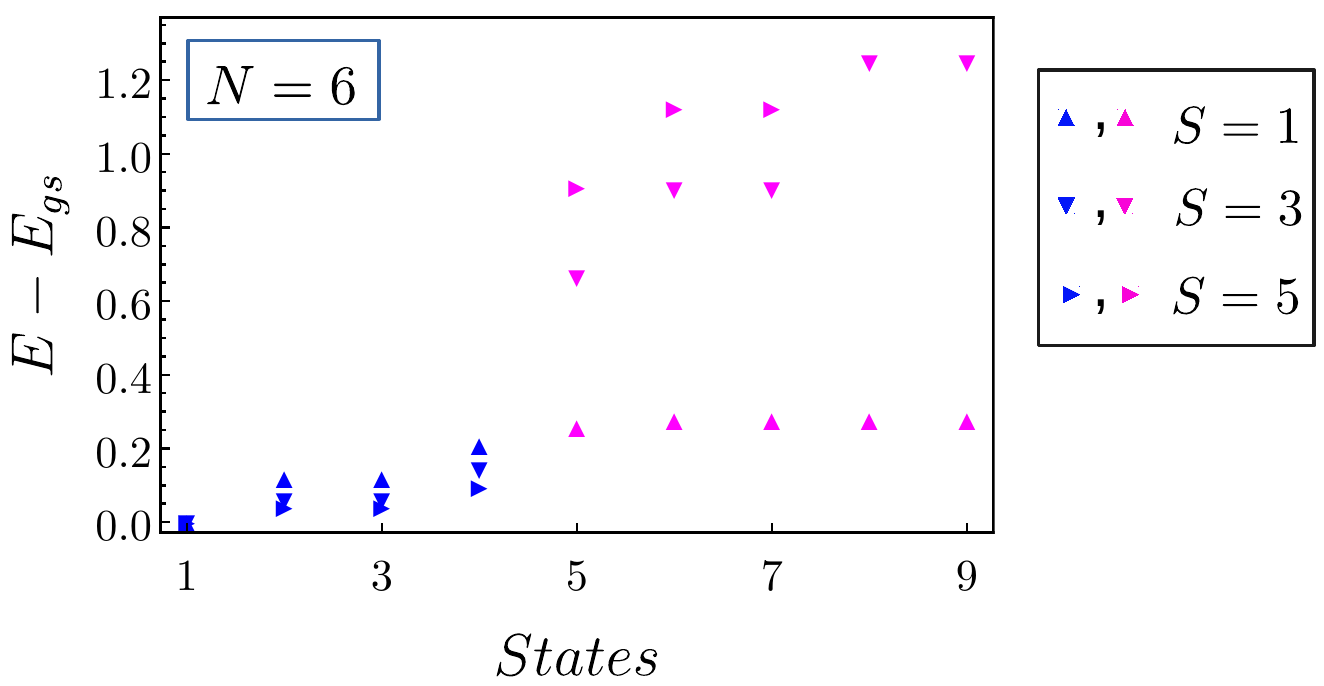}
\caption{Low-energy spectrum for the $K$-$J$ model with $N=6$, $K=1$ and $J=0.05$. We show the spectrum for $S=1,3,5$. }
\label{fig.ObS_varying_S}
\end{figure}

In Fig.~\ref{fig.ObS_varying_S}, we present the low-energy spectrum of the Kitaev - XY hexagon ($N=6$) for three different values of $S$. We show the lowest four eigenvalues using a different colour to highlight incipient localization. As $S$ increases, the system moves towards a four-fold degenerate ground state. The lowest four states becomes progressively separated from higher-energy states. We interpret this as localization due to ObP. As the fluctuation-generated potential has four distinct minima, we have four harmonic wells at low energies. The four-fold ground-state represents distinct localized states at each minimum. 

For finite $S$ values, the degeneracy of the four-fold ground-state is lost due to tunneling processes. To see this, we expand on the arguments in Sec.~\ref{ssec.well} above in the context of the energy scale $E_1$. Assuming that we have four deep harmonic wells, we obtain four Gaussian ground states -- one for each well. These states have an inherent length scale given by $\xi \sim \sqrt{\frac{\hbar}{\mu\omega}}\sim \sqrt{\frac{1}{N\sqrt{S}}}$. For small values of $N$ and $S$, this length scale can be large enough to give rise to substantial overlaps between neighboring wells. This breaks the four-fold degeneracy of the ground state. This is consistent with Fig.~\ref{fig.ObS_varying_S} where the spread in the low-energy four-fold set decreases with increasing $S$. 

\begin{figure}
\includegraphics[width=3in]{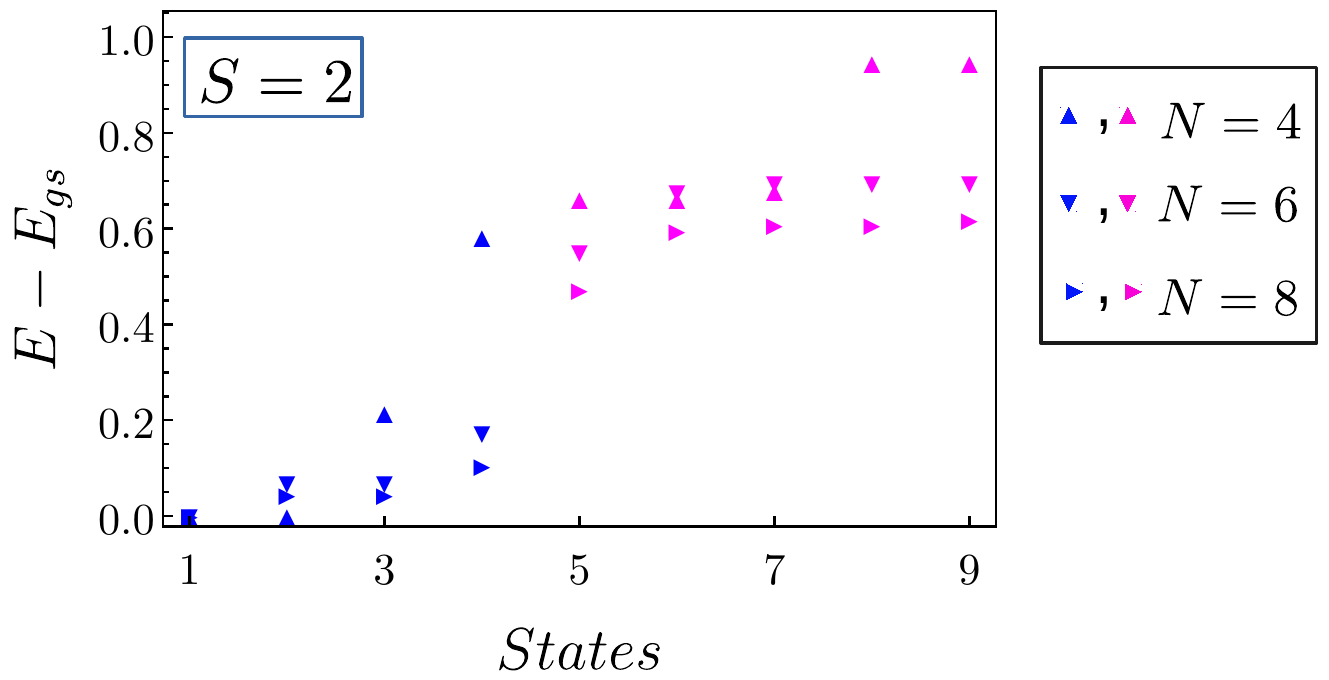}
\caption{Low-energy spectrum for the $K$-$J$ model with $S=2$, $K=1$ and $J=0.05$. We show the spectrum for $N=4,6,8$. }
\label{fig.ObS_varying_N}
\end{figure}

In Fig.~\ref{fig.ObS_varying_N}, we show the low-energy spectrum with increasing system size. Keeping $S$ fixed at $2$, we consider $N=4,6,8$. The low-energy spectrum shows a clear approach to localization with increasing $N$. We have a four-fold set of low-energy states, corresponding to the four minima in the potential. Their spread decreases with increasing $N$. This is consistent with the criterion for ObP given in Eq.~\ref{eq.criterion}.

\begin{figure}
\includegraphics[width=3in]{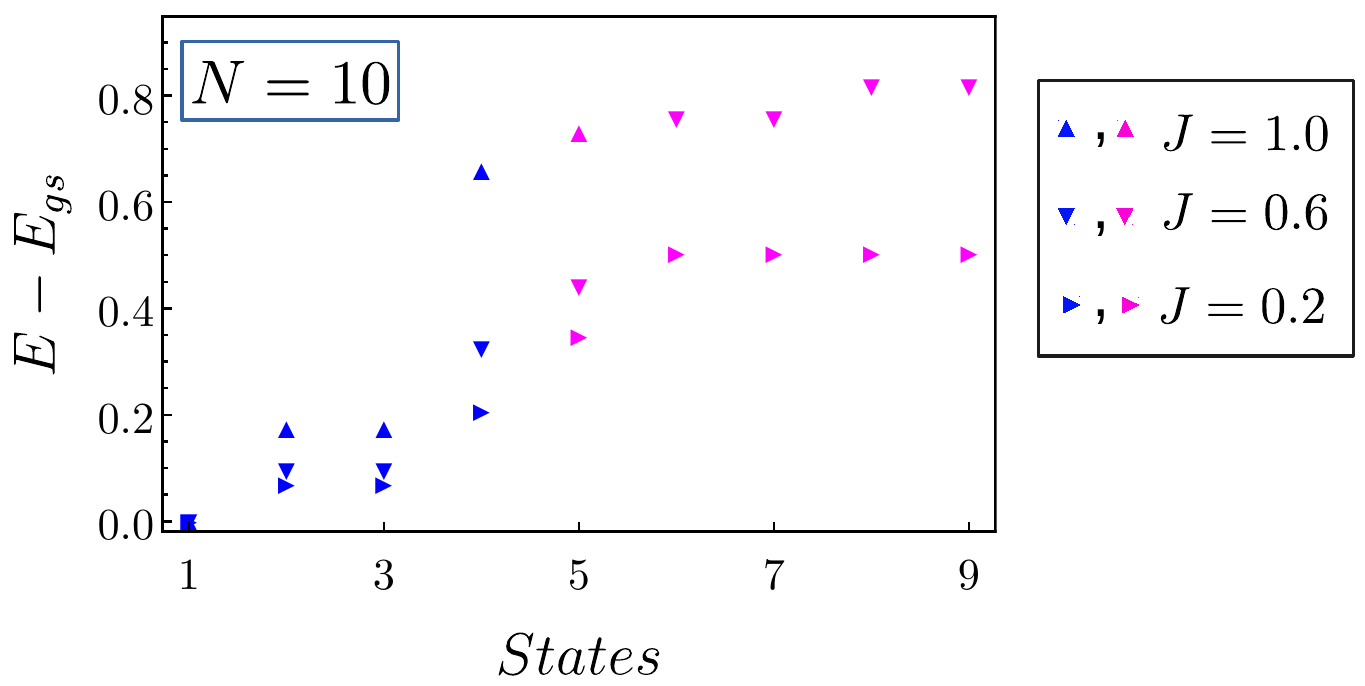}
\caption{Low-energy spectrum for the $K$-$J$ model with $N=10$, $S=1$ and $K=1$. We show the low-energy spectrum for three values of $J$.}
\label{fig.ObS_varying_J}
\end{figure}

We present a further test of the ObP paradigm in Fig.~\ref{fig.ObS_varying_J}. As shown in Fig.~\ref{fig.JK_deltaE}, the potential wells become deeper with decreasing $J/K$. The curvature of the potential about each minimum ($\gamma$ as defined in Sec.~\ref{ssec.well}) increases. This enhances the tendency of the particle to localize. This is reflected in the criterion set out in Eq.~\ref{eq.criterion}, as $\sqrt{a/\gamma}$ decreases with decreasing $J/K$. Fig.~\ref{fig.ObS_varying_J} shows the change in the low-energy spectrum with $J/K$. As $J/K$ decreases, we approach a four-fold degenerate ground state. 

\section{Distinguishing ObP and ObS}

In the $K$-$J$ model, as long as $J>0$, the potential in Fig.~\ref{fig.JK_deltaE} has four harmonic wells. We can understand localization by considering $(N\sqrt{S})$, the quantity in Eq.~\ref{eq.criterion}, as a tuning parameter. 
When this quantity is large, the particle is localized in one of the harmonic wells. As $N\sqrt{S}$ is decreased, the potential $g(\phi)$ in Eq.~\ref{eq.Lagrangian} loses significance. The physics approaches that of a free particle. This can be stated as follows: state selection by ObP can be weakened by decreasing $S$ or system size. 

\begin{figure*}
\includegraphics[width = 0.9\textwidth]{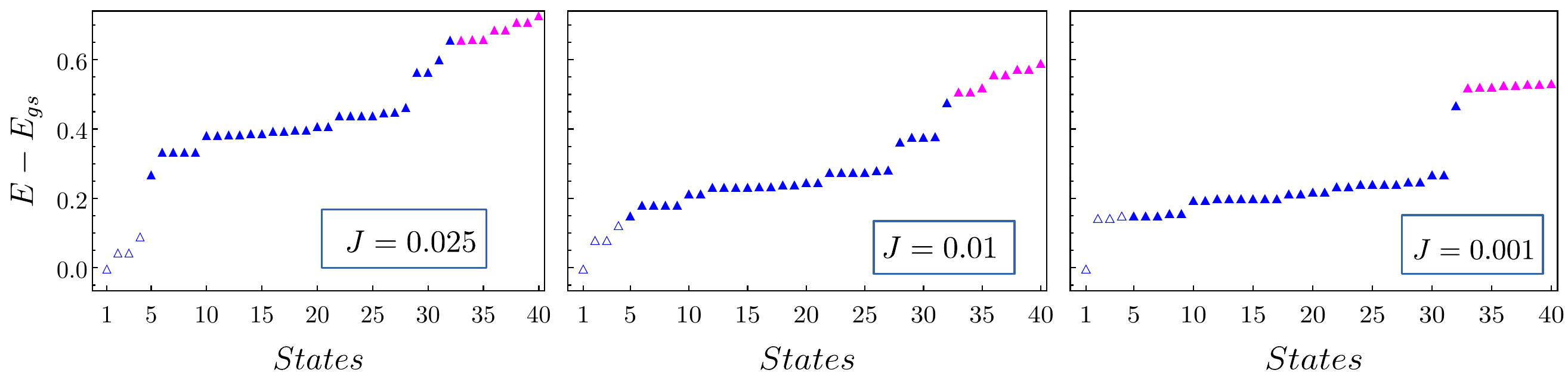}
\caption{Low-energy spectra of the $K-J$ model for small values of $J/K$. In all panels, we fix $N = 8,~S=2,~\mathrm{and}~ K= 1$. The lowest 32 (the number of Cartesian states for $N=8$) states are shown using blue markers, while higher states are shown in magenta. Within the low-lying set of 32, the lowest four states are shown using empty markers and states 5-32 are shown using filled markers. }
\label{fig.spec_tiny_J}
\end{figure*}

We compare this with state selection as $J\rightarrow 0$.  
In this limit, the potential becomes singular, resembling the $\vert x \vert$ function rather than a harmonic well. In addition, the underlying space (the CGSS) becomes larger with multiple additional pathways. Here, the free-particle problem already shows localization due to bound-state formation as described in Sec.~\ref{sec.boundstate}. Even if the potential is made weaker by decreasing $N$ or $S$, localization persists. In other words, we have pre-formed bound states due to ObS. The potential generated by quantum fluctuations merely makes them even more bound.  
As an aside, if the potential becomes very strong, it may induce additional bound states. However, in our numerics, we do not see any evidence of such extra bound states.

 We have argued that ObS selection cannot be weakened by changing system size or $S$.  
This comes with a caveat. At very low $S$, the mapping to the single-particle problem fails, leaving no room for ObS. The action in Eq.~\ref{eq.Lagrangian} is based on the spin path integral approach, which is justified in the large-$S$ limit. For very small $S$ values, the action may not serve as a good effective description.
This can be stated in terms of an energy cutoff. The magnet maps to a particle on the CGSS for energies below a certain threshold. If bound states are to be relevant, their energies must lie below the energy threshold. At the same time, we have seen that hybridization leads to a spread in the bound-state energies. If this spread is greater than the cutoff, ObS no longer provides a reliable low-energy description.  

These arguments are consistent with our numerical results. With $J > 0$, ObP weakens when $N$ or $S$ is decreased as shown in Figs.~\ref{fig.ObS_varying_S} and \ref{fig.ObS_varying_N}. When $J=0$, we find strong ObS even for $N=4$. For all $N$, we find well-separated bound states only for $S>2$. For $S=1$, we do not find a clear set of bound states as other states seem to intervene at low energies. 

An interesting regime emerges in the $K-J$ model when $J/K$ is small, with ObS and ObP competing with each other. As long as $J$ is non-zero, the CGSS is a circle -- a smooth manifold. Naively, this forbids ObS as there is no self-intersection. State selection must then occur due to ObP which will `select' precisely four states as discussed in Sec.~\ref{sec.HKJ} above. However, at the classical level, we have a large number of low-energy states with an energy cost that scales as $\sim J/K$. These form pathways that are self-intersecting and capable of hosting ObS-induced bound states. If ObS dominates, it may `select' these bound states even though they do not lie on the actual CGSS. In this case, we will have $2^{N/2 +1}$ low-lying states at the bottom of the spectrum. In Fig.~\ref{fig.spec_tiny_J}, we show numerically obtained low-energy spectra for small values of $J/K$. When $J=0.025$, ObP seems to play a dominant role as we have four well-separated states at the bottom. For lower values of $J$, these four states appear to mix with others. At the same time, a gap emerges that separates the lowest 32 states ($2^{N/2+1}$ for $N=8$) from other higher states. This indicates that ObP ceases to operate while ObS sets in, even though the CGSS is technically a smooth manifold. This suggests that ObS is a stronger mechanism for localization, atleast in this case. More broadly, ObS may operate in many materials and models that have smooth CGSS' provided there is a self-intersecting space of configurations with low energy-cost. 

\section{Discussion}
\label{sec.summary}
We have described two mechanisms for state selection in magnets with accidental degeneracy.
In each case, our analysis brings out a qualitative picture in analogy with localization. It also offers a framework to understand whether state selection will occur at all. This bears relevance to studies on spin liquids in general \cite{Zhou2017,Savary2016}, and on spiral liquids \cite{Yao2020,Gao2017} in particular. Our entire discussion is at zero temperature where fluctuations are of quantum-mechanical origin. An interesting future direction is to explore whether two distinct mechanisms exist in the case of thermal fluctuations. It is well known that thermal fluctuations can give rise to selection; in fact, `order by thermal disorder'\cite{Villain1980} predates `order by quantum disorder'\cite{Shender1982}. We note that results by Moessner and Chalker suggest that singularities play a strong role in systems with purely thermal fluctuations \cite{Moessner1998}.

Previous studies on order by disorder have used a standard prescription, selecting the ordered state with the lowest zero point energy contribution. The mechanisms discussed in this article put this prescription on firm ground. In cases where the ground-state space is a smooth manifold, the prescription simply picks the deepest minimum of a fluctuation-generated potential. If the space self-intersects, the prescription typically picks the singular point as it has additional soft modes that lower the zero point energy. An interesting future direction is to find systems where the mechanisms can compete, e.g., in the presence of multiple singularities with different co-dimensionalities. 

Our results clarify the role of quantum fluctuations at large $S$. Naively, we may expect quantum effects such as ObP and ObS to weaken and disappear with increasing $S$. However, our analysis shows that the opposite is true. The gap associated with state selection increases with S in both ObP\cite{Rau2019} ($\sim S^{1/2}$) as well as in ObS ($\sim S$). Note that these scaling relations are relevant for models studied in this article. The scaling may differ in other systems, e.g., if there are singularities with higher co-dimensionality. In order to rationalize the increase in the gap with $S$, we first define the approach to the classical limit as follows: we take $S$ to infinity while the system size and the coupling constants ($K$ and $J$) are held fixed. Note that this definition leads to a systematic increase in the bandwidth of the full problem. Nevertheless, it provides a sharp definition of the classical limit. We now consider the energy of a state as an expansion in $S$ as given by standard spin-wave theory. We have $E(S) \approx E_{cl.}(K,J) S^2 + E_{qu.}(K,J) S + \mathcal{O}(S^{1/2} )$. Here, the $\mathcal{O}(S^2)$ contribution is the classical energy while the quantum correction is $\mathcal{O}(S)$. This energy can be rewritten as $E(S) \approx S^2 \{E_{cl.}(K,J) + E_{qu.}(K,J) /S\}$, where quantum effects take the form of a $1/S$ correction. The latter form suggests that with increasing $S$, quantum effects become weaker vis-\`a-vis the classical energy. However, this is not relevant in the context of state selection. As we have multiple states that have precisely the same classical energy, it is more appropriate to view quantum fluctuations as an $\mathcal{O}(S)$ effect. This view is in consonance with the state selection gap increasing with $S$.

Our work builds on several earlier studies that suggest that quantum fluctuations generate a localizing potential \cite{Rau2019,Belorizky1980,Shender1982}. In Eq.~\ref{eq.criterion} above, we formulate a rule of thumb to determine when ObP becomes effective. This can be particulary useful for finite-sized systems such as magnetic flakes and molecular magnets\cite{Mller2000,Jian2006,Zaharko2008,Nehrkorn2010,Lin2011,Park2016}. It can also provide insight into numerical studies which are necessarily limited to finite sizes and finite spin lengths ($S$ values). The criterion in Eq.~\ref{eq.criterion} also provides an interesting contrast between ObP and ObS, as ObP requires large system sizes in order to bring about localization. In contrast, ObS does not seem to place strong constraints on system size. 
This suggests that ObS is much more effective than ObP in small systems such as molecular magnets. This is consistent with results presented in Ref.~\onlinecite{Khatua2019} contrasting the symmetric XY quadrumer and the asymmetric XY quadrumer. The former allows for ObS and in fact, shows strong state selection. In contrast, the latter has a smooth manifold as its CGSS. Despite the possibility of ObP, it does not show state selection even as $S$ is tuned to large values.

In the one-dimensional spin-$S$ Kitaev model, we have shown that a network-like structure emerges at low energies. Intriguingly, the size of the Kitaev spin chain tunes the complexity of the network. In particular, increasing the system size increases the number of wires that cross at each node. This is of interest to the theory of quantum graphs that discusses solutions of Schr\"odinger-like equations on networks\cite{Pauling1936,Kottos1997,Keating2008,
Exner2008,Harrison2011,Alexandradinata2018}. The Kitaev spin chain offers an interesting test case with tunable complexity.

A crucial question is whether ObS survives in the thermodynamic limit. We are unable to definitively demonstrate this numerically. We hope further studies will clarify this question. The phenomenon of ObS will add a new dimension to studies of Kitaev-like models that have hitherto used traditional spin-wave-based methods\cite{Yang2020}. Many studies have focused on the spin-1/2 Kitaev model \cite{Feng2007,Nussinov2009,Fan2018,Kitaev2006}, using a mapping to Majorana fermions. Such fermionization approaches do not generalize to $S>1/2$. Our results motivate a deeper look into suitable effective pictures for $S>1/2$, given that larger values of $S$ are conducive to ObS due to bound-state formation. This question may soon acquire experimental relevance with several proposals for realizing Kitaev systems with $S>1/2$ \cite{Koga2018,Suzuki2018,Oitmaa2018,Minakawa2019,Stavropoulos2019}.
Our discussion of ObS may have relevance beyond the one-dimensional spin-$S$ Kitaev model. For example, Ref.~\onlinecite{Chandra2010} contains hints that the spin-S Kitaev model on the honeycomb lattice may have a self-intersecting CGSS.

\acknowledgements S.K. acknowledges support from the Shastri Indo-Canadian Institute as part of the Shastri Research Student Fellowship Programme (SRSF 2019). S.K. also thanks the group of Prof. Michel Gingras at the University of Waterloo for gracious hospitality. We thank Kristian Tyn-Kai Chung for useful inputs.

\appendix
\section{Energy minimization and the CGSS }
\label{app.CGSS_completeness}
We follow the approach of BSS~\cite{Baskaran2008} to minimize energy in the one-dimensional spin-$S$ Kitaev model of Eq.~\ref{eq.H_Kitaev}. Treating spins as classical vectors, we have three components per spin. On a chain with $N$ sites, we have $3N$ independent variables. However, they are constrained to maintain the length of each spin fixed at $S$. We use the method of Lagrange multipliers to enforce these constraints, defining
\begin{eqnarray}
H_\lambda = \frac{K}{2}\sum_{j = 0}^{N-1} \lambda_j\left[(S_j^x)^2 + (S_j^y)^2 + (S_j^z)^2 - S^2  \right]. 
\end{eqnarray}
We minimize the energy using the conditions, $\partial(H_K - H_\lambda)/\partial S_i^\alpha = 0, $ where $S_i^\alpha$ is the $\alpha$-component ($\alpha=x,y,z$) of the spin at site $i$. A detailed discussion, specialized to the case of a 4-site chain, can be found in Appendix~A of Ref.~\onlinecite{Sarvesh2020}. These arguments readily generalize to a chain of arbitrary length. They lead to the conclusion that $\lambda_j = -1$ for every $j$. This further leads to the following two conditions. On every site, the z-component of the spin must vanish, i.e., $S_i^z=0$ for every $i$. Secondly, if sites $i$ and $i+1$ are connected by an $x-x$ bond, we must have $S_i^x = -S_{i+1}^x$. However, if sites $i$ and $i+1$ are connected by a $y-y$ bond, we have $S_i^y = -S_{i+1}^y$.
With these conditions, we arrive at the ground-state energy, $E_{min} = -NKS^2/2$.

As described in the main text, Cartesian states immediately satisfy these conditions. However, they are not the only ground states. To show this, we first consider an arbitrary Cartesian state of the $x$ family. We describe this state by specifying the spin vector at each site, labeling the sites as $i=0,1,2\ldots,N-1$ where $i=0$ and $i=N$ are taken to be identical on account of periodic boundary conditions.
\begin{widetext}
\begin{eqnarray}
\label{eq.xcartesian}
{X_{\sigma_1 \sigma_2 \cdots}} &\equiv& \{S_0 = (S \sigma_1,0), \phantom{a}S_1 = (-S\sigma_1,0), \phantom{a}S_2 = (S\sigma_2,0), \phantom{a}S_3 = (-S\sigma_2,0), \phantom{a}S_4 = (S\sigma_3,0), \phantom{a} S_5 = (-S\sigma_3,0),\cdots\},\nonumber \\
\label{eq.Xcartesian}
\end{eqnarray}
We only specify the $x$ and $y$ components of each spin as the $z$ component is always zero. Here, $\sigma_1,~\sigma_2,\ldots,~\sigma_{N/2}$ are Ising variables that define the $X$-Cartesian state, with each $\sigma$ being $\pm 1$. To understand the energy content of this state, we define `bond energy', $E_{i,i+1}$, as the energy contribution from the bond connecting sites $i$ and $i+1$. We have
\begin{eqnarray}
E_{01}=-KS^2, ~E_{12} = 0, ~E_{23}=-KS^2, ~E_{34}=0,~E_{45}=-KS^2,\ldots.
\end{eqnarray}
We see that each $x-x$ bond offers the same negative contribution to the ground-state energy. The $y-y$ bonds do not contribute. This is consistent with the expression for ground-state energy ($E_{min}$) given above. 
We next consider a generic Cartesian state of the $y$ family,
\begin{eqnarray}
{Y_{\mu_1 \mu_2 \cdots}} &\equiv& \{S_0 = (0,-S\mu_{N/2}), \phantom{a}S_1 = (0,S\mu_1), \phantom{a}S_2 = (0,-S\mu_1), \phantom{a}S_3 = (0,S\mu_2), \phantom{a}S_4 = (0,-S\mu_2), \phantom{a} S_5 = (0,S\mu_3),\cdots\}.\nonumber\\
\end{eqnarray}
We denote the Ising moments here as $\mu_1,~\mu_2,\ldots,~\mu_{N/2}$, with each $\mu$ being $\pm 1$. The bond energies in this state are given by
\begin{eqnarray}
E_{01}=0, ~E_{12} = -KS^2, ~E_{23}=0, ~E_{34}=-KS^2,~E_{45}=0,\ldots.
\end{eqnarray}
Here, every $y-y$ bond offers the same negative contribution to the ground-state energy while the $x-x$ bonds do not contribute.
We now present a smooth energy-preserving transformation that connects these two Cartesian states. We use a single parameter $\phi \in [0,\pi/2]$ to define a configuration,
\begin{eqnarray}
\label{eq.transformation}
\{ S_0 = S(\sigma_1 c_\phi,-\mu_{N/2} s_\phi), \phantom{a}S_1 = S(-\sigma_1 c_\phi,\mu_1 s_\phi), \phantom{a}S_2 = S(\sigma_2 c_\phi,-\mu_1 s_\phi), \phantom{a}S_3 = S(-\sigma_2 c_\phi,\mu_2 s_\phi),\nonumber\\ \phantom{a}S_4 = S(\sigma_3 c_\phi,-\mu_2 s_\phi), \phantom{a} S_5 = S(-\sigma_3 c_\phi,\mu_3 s_\phi) ,\cdots\}, 
\end{eqnarray}
where $c_\phi\equiv \cos\phi, ~s_\phi\equiv\sin\phi$. 
This configuration is designed such that it reduces to ${X_{\sigma_1 \sigma_2 \cdots}} $ when $\phi = 0$ and to  ${Y_{\mu_1 \mu_2 \cdots}}$ at $\phi=\pi/2$. In the above expressions, we have explicitly written out the forms of the first few spins. Indeed, all spins can be written in an analogous fashion. We note that at intermediate values of $\phi$, the spins have non-zero components along both $x$ and $y$ axes. We now examine the bond energies in this configuration, 
\begin{eqnarray}
E_{01}=-K S^2 c_\phi^2, ~E_{12} = -KS^2 s_\phi^2, ~E_{23}=-K S^2 c_\phi^2, ~E_{34}=-KS^2 s_\phi^2,~E_{45}=-K S^2 c_\phi^2,\ldots.
\end{eqnarray}
All bonds contribute to the ground-state energy. Each pair of adjacent bonds contributes $-KS^2 (c_\phi^2 + s_\phi^2 ) = -KS^2$. The overall ground-state energy remains constant as $\phi$ is varied. 
\end{widetext}
We have demonstrated that any pair of Cartesian states of the form $({X_{\sigma_1 \sigma_2 \cdots}},{Y_{\mu_1 \mu_2 \cdots}} )$ is smoothly connected by a one-parameter family of states. Note that no such transformation exists for two Cartesian states that belong to the same family (i.e., two states constructed from the same underlying dimer cover). However, they are connected indirectly. That is, we can smoothly go from one $X$-Cartesian state to another via an intermediate $Y$-Cartesian state. This picture leads to the network-like CGSS depicted in Fig.~\ref{fig.wires}

We have argued that Cartesian states readily satisfy the ground-state conditions. We have also demonstrated that each inter-family pair of Cartesian states is connected by a one-parameter family of states. We next argue that these considerations exhaust all possible ground states. While a general proof is not possible, we will show below that the CGSS, as described, is a closed space. That is, in the vicinity of any point on our network-like CGSS, the only states that satisfy the energy minimization conditions are those on the network itself. We show this in two steps: (i) We first consider a generic element of the CGSS, corresponding to an intermediate point on a segment that connects two nodes. We consider all possible small deviations from this state. If we are to satisfy the energy minimization conditions, we may only allow changes in one coordinate, i.e., we have only  one soft mode. (ii) We next consider a node and enumerate all possible small deviations about the corresponding Cartesian state. We find precisely $N_c/2$ soft modes, where $N_c$ is as defined in Sec.~\ref{sec.model} of the main text. This can be interpreted as $N_c$ line segments emanating from the node -- precisely as conceived in our description. These arguments show that our network-like description of the CGSS is consistent.

We consider a generic element of our CGSS as given in Eq.~\ref{eq.transformation}, with $\phi$ being neither zero nor a multiple of $\pi/2$. We consider all possible (small) deformations about this state. As we have three-component spins with fixed lengths, we have $2N$ degrees of freedom where $N$ is the number of sites. As energy minimization requires the z-component of each spin to be zero, we may simply neglect fluctuations that take the spins out of the plane. This leaves us with $N$ degrees of freedom. Accordingly, we introduce one angle variable, $\delta_i$, for each site,
\begin{eqnarray}
\label{eq.deform}
\{ S_0 &=& S(\sigma_1 c_{\phi+\delta_0} ,-\mu_{N/2} s_{\phi+\delta_0}), \nonumber \\
\phantom{a}S_1 &=& S(-\sigma_1 c_{\phi+\delta_1},\mu_1 s_{\phi+\delta_1}), \nonumber \\
\phantom{a}S_2 &=& S(\sigma_2 c_{\phi+\delta_2},-\mu_1 s_{\phi+\delta_2}),  \nonumber \\
 \phantom{a}S_3 &=& S(-\sigma_2 c_{\phi+\delta_3},\mu_2 s_{\phi+\delta_3}),\cdots\}. 
\end{eqnarray}
As we are interested in small deviations, we assume that the $\delta_i$'s are small.  
We now demand that the fluctuations must satisfy the energy minimization conditions. This leads to 
\begin{eqnarray}
c_{\phi+\delta_0} = c_{\phi+\delta_1}, ~s_{\phi+\delta_1} = s_{\phi+\delta_2}, ~c_{\phi+\delta_2} = c_{\phi+\delta_3}, \cdots.
\end{eqnarray}
In order to satisfy these equations, all $\delta$'s must be equal. We are left with a one-parameter deformation that preserves the ground-state energy. All other deviations take us out of the CGSS. This can be restated as follows: in the vicinity of a generic point, the CGSS is one-dimensional. 

We next consider a Cartesian state. For concreteness, we take a generic state of the x-family as given in Eq.~\ref{eq.Xcartesian}. The arguments extend to Cartesian states of y-family as well. We are interested in deformations about this state that preserve the ground-state energy. As the minimization conditions require all spins to lie in the XY plane, we neglect out-of-plane deformations to write 
\begin{eqnarray}
\{S_0 &=& S(\sigma_1 c_{\delta_0},s_{\delta_0}), S_1 = 
 S(-\sigma_1 c_{\delta_1},s_{\delta_1}), \nonumber \\
 S_2 &=&  S(\sigma_2 c_{\delta_2},s_{\delta_2}),S_3 = 
 S(-\sigma_2 c_{\delta_3},s_{\delta_3}),\nonumber  \\
 S_4 &=& 
 S(\sigma_3 c_{\delta_4},s_{\delta_4}),S_5 = 
 S(-\sigma_3 c_{\delta_5},s_{\delta_5}),\ldots    \}.~~~
\end{eqnarray} 
We have introduced an angle variable, $\delta_i$, for every site $i$. We denote $c_{\delta_i} \equiv \cos \delta_i$ and $s_{\delta_i} \equiv \sin\delta_i$. As we are only interested in small deviations from the Cartesian state, we assume that the $\delta$'s are small. We now demand that the deformed state must satisfy the mnimization conditions set out above, leading to 
\begin{eqnarray}
c_{\delta_0} =c_{\delta_1};~c_{\delta_2} =c_{\delta_3};~\ldots, \nonumber \\
s_{\delta_1}=-s_{\delta_2}; ~s_{\delta_3}=-s_{\delta_4}; ~\ldots.
\end{eqnarray}
To satisfy the constraints in the second line, we must have $\delta_1 = -\delta_2, ~\delta_3 = -\delta_4,~$ etc. 
This halves the number of degrees of freedom. 
We next consider the constraints given in the first line. They give rise to $\delta_0 = \pm \delta_1, ~\delta_2 = \pm \delta_3,~$ etc. Put together, they constrain all $\delta$'s to have the same amplitude. However, they may differ in sign. We arrive at 
\begin{eqnarray}
\{ 
\delta_1,\delta_2,\delta_3,\delta_4,\ldots
\}=\{\xi_1 \delta, -\xi_1 \delta, \xi_2 \delta, -\xi_2 \delta,\ldots  
\}.~~~
\end{eqnarray}
We have introduced Ising-like variables with each $\xi$ taking the value $\pm 1$. With $N/2$ free Ising variables, we have $2^{N/2}$ deformations that preserve the ground-state energy. This reveals the geometry of the CGSS within configuration space. Each Cartesian state is a node that has $2^{N/2}$ wires emanating from it. This is consistent with the picture of the CGSS described in Sec.~\ref{sec.model} and depicted in Fig.~\ref{fig.wires}.

\section{The CGSS of the $K$-$J$ model and the role of quantum flucuations}
\label{app.sw_JK}
We rewrite the $K$-$J$ Hamiltonian of Eq.~\ref{eq.HKJ} as 
\begin{eqnarray}
H_{K-J} = \sum_{j=1}^{N/2}\left[ J \left( S^x_{j,A}S^x_{j,B} + S^y_{j,A}S^y_{j,B} + S^x_{j,B}S^x_{j+1,A}+\right.\right.\nonumber\\
\left.\left. S^y_{j,B}S^y_{j+1,A}  \right) + K \left( S^x_{j,A}S^x_{j,B} + S^y_{j,B}S^y_{j+1,A} \right)\right].~~~
\end{eqnarray}
We have assumed a two-site unit cell, with sub-lattices labeled as $A$ and $B$. As described in Sec.~\ref{sec.HKJ} of the main text, the CGSS is a circle parametrized by an angle variable, $\phi$. We define an element of the CGSS using    
\begin{eqnarray}
		\vec{S}_{j,A} = S\hat{n},\hspace{3mm} \vec{S}_{j,B} = -S\hat{n},\hspace{3mm} \hat{n} = \cos\phi \,\hat{x} + \sin\phi\, \hat{y}.
\end{eqnarray}
Following BSS, we perform a Holstein-Primakoff analysis by defining 
\begin{eqnarray}
		\vec{S}_{j,A} &= S \left( 1 - \frac{q_{j,A}^2+p_{j,A}^2}{2S}\right) \hat{n} + \sqrt{S} \left(q_{j,A}\hat{e} + p_{j,A} \hat{z} \right),\nonumber \\
	\vec{S}_{j,B} &= - S \left( 1 - \frac{q_{j,B}^2+p_{j,B}^2}{2S}\right) \hat{n} - \sqrt{S} \left(q_{j,B}\hat{e} + p_{j,B} \hat{z} \right),\nonumber\\ 
\end{eqnarray}
where $\hat{e} = -\sin\phi \,\hat{x}+\cos\phi \,\hat{y}$ is the vector orthonormal to $\hat{n}$ in the XY-plane. The $p$ and $q$ variables are canonically conjugate with $\comm{q_{j,\alpha}}{p_{l,\beta}} = i\delta_{jl}\delta_{\alpha \beta}$, where $\alpha,\beta = A,B$. In terms of these coordinates, the spin-wave Hamiltonian takes the following form in momentum space,
\begin{align}
H^{\mbox{sw}}(\phi) = &(2J + K)S\sum_{k=0}^\pi\left[ \begin{pmatrix}
p_{-k,A} & p_{-k,B}
\end{pmatrix}
\begin{pmatrix}
1 & 0\\
0 & 1
\end{pmatrix}
\begin{pmatrix}
p_{k,A} \\
p_{k,B}
\end{pmatrix}\right.\nonumber\\
&\left. + 
\begin{pmatrix}
q_{-k,A} & q_{-k,B}
\end{pmatrix}
\begin{pmatrix}
1 & f_k(\phi)\\
f^*_k(\phi) & 1
\end{pmatrix}
\begin{pmatrix}
q_{k,A} \\
q_{k,B}
\end{pmatrix}
\right],\nonumber\\
\end{align}
where $f_k(\phi) = -\frac{1}{2J+K} (J + K\sin^2\phi + (J+K\cos^2\phi)e^{ik})$. Diagonalizing this Hamiltonian, we obtain the spin-wave energies as $S\epsilon_\phi (k,\alpha)$ where
\begin{eqnarray}
		\epsilon_\phi (k,\alpha) = (2J+K)\sqrt{1+\alpha \abs{f_k(\phi)}}.
\end{eqnarray}
Here, we have two bands indicated by an index $\alpha =\pm 1$. 
The total zero point energy is given by 
\begin{align}
E^{\mbox{sw}}(\phi) = (2J + K)S\sum_{k = 0}^\pi\left( \sqrt{1 + \abs{f_k(\phi)}} + \sqrt{1 - \abs{f_k(\phi)}} \,\right).\nonumber\\
\label{eq.zpe}
\end{align} 
This is precisely the quantity defined as $g(\phi)$ in Eq.~\ref{eq.gphi} of the main text and plotted in Fig.~\ref{fig.JK_deltaE}.
To minimize this quantity, we first note that $f(x) = \sqrt{1+\abs{x}} + \sqrt{1-\abs{x}}$ is a monotonically decreasing function of $\abs{x}$. The zero point energy depends on the ground-state parameter $\phi$ via $f_k(\phi)$. 
To see this, we write
\begin{eqnarray}
\abs{f_k(\phi)}^2 = \frac{1}{(2J+K)^2}\left(4J(J+K)\cos^2(k/2)\right.\nonumber\\
\left. + K^2(1-\sin^2(k/2)\sin^2(2\phi))\right).
\end{eqnarray}
For any given $k$, we see that this quantity is maximum when $\sin^2(2\phi)=0$, i.e., when $\phi = 0,\pi/2,\pi,3\pi/2$. Note that these choices correspond to Cartesian states. It follows that, at each $k$, the zero point energy contribution is minimum when $\abs{f_k(\phi)}$ is maximum, i.e., when $\phi$ takes one of the four values given above. 

\bibliographystyle{apsrev4-1} 
\bibliography{1D_Kitaev.bib}
\end{document}